\documentclass[%
 reprint,
 amsmath,amssymb,
 aps,
 prapplied,
]{revtex4-2}

\usepackage{graphicx}
\usepackage{dcolumn}
\usepackage{bm}

\begin{document}

\preprint{APS/123-QED}

\title{Data-driven signal--idler discrimination for faster qubit readout with a traveling-wave parametric amplifier}

\author{Javier Navarro Montilla}
\author{Aneirin J. Baker}
\email{Currently at Rigetti Computing}
\author{Narendra Acharya}
\author{Vivek Chidambaram}
\email{Email: vivek.chidambaram@stfc.ac.uk}
 \affiliation{National Quantum Computing Centre (NQCC), Rutherford Appleton Laboratory, Harwell Campus, Didcot, Oxfordshire, OX11 0QX, United Kingdom}

\date{\today}

\begin{abstract}

Superconducting parametric amplifiers enable fast, high-fidelity qubit readout through near-quantum-limited amplification. In phase-preserving parametric amplification, amplification of the signal is accompanied by the generation of a correlated idler at a different frequency, as required by energy conservation. Although the idler carries qubit-state information, it is typically discarded. Here, we demonstrate qubit-state discrimination using both the signal and idler generated by a Josephson traveling-wave parametric amplifier (TWPA). We simultaneously acquire the two output modes and combine their measured quadratures using four-dimensional linear discriminant analysis (LDA), which  determines the relative signal--idler weights directly from the measured state centroids and covariance. Under the same acquisition protocol, joint signal--idler discrimination reaches the plateau fidelity of $F=0.973$ at $T_{\mathrm{int}}\approx1.39\,\mu\mathrm{s}$, compared with $1.71\,\mu\mathrm{s}$ for signal-only discrimination, corresponding to an approximately $19\,\%$ reduction in the required integration time. These results establish joint signal--idler LDA as a simple data-driven approach for exploiting information that is otherwise discarded, which reduces the integration time required for superconducting-qubit readout.

\end{abstract}

\maketitle

\section{Introduction}
\label{sec:sec1_intro}

Superconducting circuits have emerged as a promising platform for fault-tolerant quantum computing, supported by advances in qubit coherence, control, and high-fidelity gate operations~\cite{kjaergaard2020superconducting}. Improvements in readout technology, particularly quantum-limited amplification, have also made it possible to measure qubit states with increasing speed and accuracy~\cite{Walter2017Rapid, Spring2025Fast}. As processors grow and quantum error correction demands repeated mid-circuit measurements, readout must deliver high fidelity in short measurement times while efficiently using available hardware~\cite{krinner2022Realizing, Acharya2025Quantum}.

Superconducting qubits are typically measured using dispersive readout, where the qubit state is inferred from the frequency shift of a readout resonator coupled to the qubit~\cite{blais2004cavity, wallraff2004strong}. This shift is detected through the amplitude and phase of a microwave probe tone. By assigning different resonance frequencies to different readout resonators, multiple qubits can be simultaneously measured on a single readout line. Therefore, traveling-wave parametric amplifiers (TWPAs) are particularly well suited to this architecture, combining broadband operation, gains exceeding 20\,dB, and added noise approaching the quantum limit~\cite{Macklin2015Anear, planat2020photonic,  Gaydamachenko2025rfsquid, ranadive2022kerr}.

TWPAs achieve gain through nonlinear wave mixing. As a weak signal at angular frequency $\omega_s$ co-propagates with a strong pump at $\omega_p$, the nonlinear interaction transfers energy from the pump into the signal and generates an idler mode. Energy conservation sets the idler frequency to $\omega_i=2\omega_p-\omega_s$ for four-wave mixing and $\omega_i=\omega_p-\omega_s$ for three-wave mixing, while efficient amplification additionally requires satisfying the phase-matching condition \cite{obrien2014resonant}. The signal and idler output fields are correlated by the parametric interaction. Although conventional dispersive readout uses only the amplified signal to discriminate the qubit state, the generated idler also acquires a state-dependent coherent displacement.

These signal--idler correlations are a fundamental feature of phase-preserving nondegenerate parametric amplification and underlie two-mode squeezing, for which appropriate joint quadratures of the two output modes exhibit reduced fluctuations below the vacuum-noise level \cite{caves1982quantum}. Such correlations have therefore been exploited in a variety of squeezing-assisted qubit-readout schemes. Examples include parametric-amplifier interferometry~\cite{barzanjeh2014interferometry}, two-mode squeezed illumination of coupled measurement cavities~\cite{didier2015heisenberg}, and squeezing generated directly within the readout cavities~\cite{govia2017locally}. Other proposals combine injected and intracavity squeezing to enhance dispersive readout~\cite{qin2024exponentially}, while stroboscopic coupling has enabled an experimental improvement in measurement sensitivity using squeezed illumination~\cite{eddins2018stroboscopic}.

Although these approaches can provide substantial improvements in readout performance, their implementation often requires additional microwave components, tailored measurement protocols, or modifications to the conventional dispersive-readout chip architecture. In contrast, the authors in Ref.~\cite{Abdo2026enhancing} recently demonstrated, using a Josephson mixer, that the signal--idler correlations generated by a phase-preserving amplifier can be exploited directly through simultaneous detection and coherent combination of the two output modes. This approach does not require an externally squeezed measurement field or modification of the qubit--resonator interaction and is compatible with frequency-multiplexed readout, providing a practical route to enhanced readout by retaining and jointly processing the idler measurement.

In this work, we demonstrate qubit-state discrimination using the idler generated by a TWPA, both independently and jointly with the amplified signal. We introduce a four-dimensional linear discriminant analysis (LDA) framework for the joint signal--idler record, and we use it to construct a theoretical model illustrating the discrimination advantage. Because LDA learns the relative quadrature weights directly from the measured centroids and covariance, it provides a data-driven alternative to prescribed coherent signal--idler combinations. We experimentally demonstrate adaptation of the learned LDA weights to changes in TWPA pump phase, and show that retaining the idler can reduce the integration time required for qubit readout.

\section{Signal--idler readout and linear discriminant analysis}
\label{sec:sec2_methods}

\begin{figure*}[t]
    \centering
    \includegraphics[width=\textwidth]{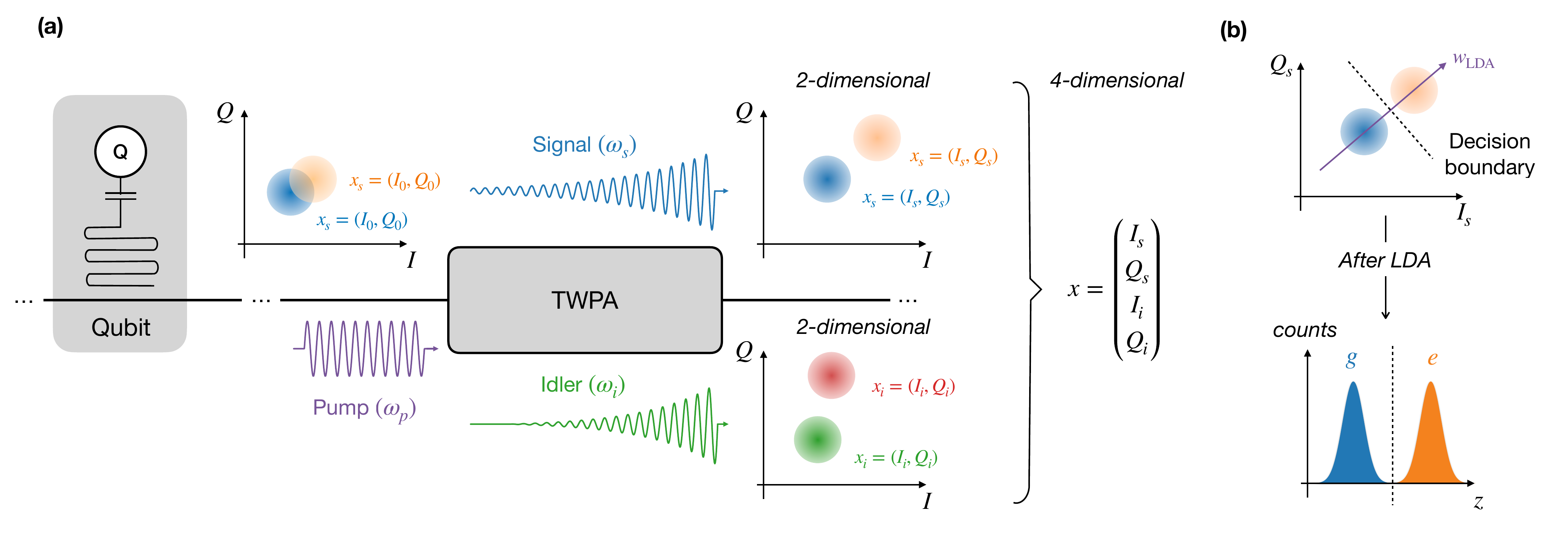}
    \caption{
    Schematic of signal--idler readout and linear discriminant analysis (LDA). (a) Superconducting-qubit dispersive readout with a TWPA as the first-stage amplifier. The qubit-state-dependent signal at frequency $\omega_s$ is amplified while an idler at frequency $\omega_i$ is generated. The simultaneously acquired signal and idler IQ measurements form a four-dimensional vector. (b) Illustration of signal-only LDA. The two-dimensional IQ data are projected onto the learned discrimination axis $w_{\mathrm{LDA}}$, which maximizes the separation of the class means relative to their within-class variance. The projected distributions are used to assign each measurement to the ground or excited state; for equal class priors and a common covariance, the threshold lies midway between the projected means.
    }
    \label{fig:dispersive_readout}
\end{figure*}

\subsection{Two- and four-dimensional linear discrimination}

Figure~\ref{fig:dispersive_readout}(a) illustrates dispersive readout with a TWPA as the first-stage amplifier. A microwave probe at angular frequency $\omega_s$ interacts with the readout resonator, whose response depends on the qubit state. The resulting state-dependent changes in amplitude and phase are measured through the in-phase ($I$) and quadrature ($Q$) components of the output field. Repeating the measurement for each prepared qubit state $q\in\{g,e\}$, corresponding to the ground and excited states, respectively, produces two distributions in the signal IQ plane. We approximate them as two-dimensional Gaussian distributions,
\begin{equation}
    x_s^{(q)} =
    \begin{pmatrix}
        I_s^{(q)}\\
        Q_s^{(q)}
    \end{pmatrix},
    \qquad  
    x_s^{(q)}\sim\mathcal{N}\!\left(\mu_s^{(q)},\Sigma_s\right).
    \label{eq:signal_lda_gaussian_model}
\end{equation}
Here $\mu_s^{(q)}$ is the state-dependent mean and $\Sigma_s$ is the common covariance matrix.

The TWPA also generates an idler field at angular frequency $\omega_i$. The parametric interaction maps the state-dependent coherent displacement of the input signal onto both output modes while generating correlations between their noise fluctuations. Joint detection gives a four-dimensional representation of each single-shot measurement,
\begin{equation}
    x^{(q)} =
    \begin{pmatrix}
        I_s^{(q)}\\
        Q_s^{(q)}\\
        I_i^{(q)}\\
        Q_i^{(q)}
    \end{pmatrix},
    \qquad
    x^{(q)}\sim\mathcal{N}\!\left(\mu^{(q)},\Sigma\right),
    \label{eq:4D_lda_gaussian_model}
\end{equation}
where $\Sigma$ describes both the noise within each mode and the correlations between them.

We use LDA to assign each outcome to a qubit state~\cite{mclachlan1992discriminant}. For two Gaussian classes with a common covariance, the discriminant direction and projected coordinate are
\begin{equation}
    w_{\mathrm{LDA}}=\Sigma^{-1}\Delta\mu,
    \qquad
    z=w_{\mathrm{LDA}}^T x,
    \label{eq:lda_direction}
\end{equation}
where $\Delta\mu=\mu^{(e)}-\mu^{(g)}$. 

For equal class priors and equal misclassification costs, the decision threshold is the midpoint between the projected class means. For the joint signal--idler measurement,
\begin{equation}
\begin{aligned}
    w_{\mathrm{LDA}} &=
    \begin{pmatrix}
        w_{s,I}\\
        w_{s,Q}\\
        w_{i,I}\\
        w_{i,Q}
    \end{pmatrix},
    \qquad
    z =
    \begin{aligned}[t]
        &w_{s,I}I_s+w_{s,Q}Q_s\\
        &+w_{i,I}I_i+w_{i,Q}Q_i
    \end{aligned}
\end{aligned}
\label{eq:lda_weight_components}
\end{equation}
The relative magnitudes and signs of these coefficients determine how the measured quadratures are combined; their overall normalization does not affect the decision boundary. While Fig.~\ref{fig:dispersive_readout}(b) illustrates signal-only LDA, the same procedure applies to idler-only and joint signal--idler discrimination. In the two-dimensional cases, the discriminant vector contains the two weights associated with the corresponding IQ plane, whereas joint 4D LDA uses the four coefficients in Eq.~\eqref{eq:lda_weight_components}.

For the common-covariance Gaussian model, the optimal projected separation is given by the squared Mahalanobis distance,
\begin{equation}
    R=\Delta\mu^T\Sigma^{-1}\Delta\mu.
    \label{eq:lda_mahalanobis}
\end{equation}
For signal-only readout with isotropic noise, $\Sigma_s=\sigma_s^2\mathbb{I}_2$, where $\sigma_s^2$ is the variance of either quadrature, this reduces to
\begin{equation}
    R_s = \frac{\|\Delta\mu_s\|^2}{\sigma_s^2} = \frac{(\Delta I_s)^2+(\Delta Q_s)^2}{\sigma_s^2}\equiv \mathrm{SNR}^2,
    \label{eq:signal_snr}
\end{equation}
where we use the amplitude-SNR convention $\mathrm{SNR}=|\Delta\mu_s|/\sigma_s$ for the separation of two Gaussian readout distributions \cite{blais2021circuit}. Thus, in the idealized isotropic Gaussian case, $R$ is simply the squared signal-to-noise ratio. Equation~\eqref{eq:lda_mahalanobis} generalizes this quantity to anisotropic and correlated noise and to higher-dimensional measurement records, such as the joint signal--idler vector of Eq.~\eqref{eq:4D_lda_gaussian_model}.

For equal priors and a common Gaussian covariance, the predicted assignment fidelity is given by~\cite{Magesan2015Machine}
\begin{equation}
    F_{\mathrm G}(R)=\frac{1}{2} \left[1+\operatorname{erf}\!\left(\sqrt{\frac{R}{8}}\right)\right].
    \label{eq:lda_gaussian_fidelity}
\end{equation}
Thus, a larger $R$ implies higher fidelity within the Gaussian model. Experimentally, however, non-Gaussian tails, state-preparation errors, and state transitions during measurement can cause the observed assignment fidelity to saturate even while $R$ continues to increase.

\subsection{Analytical prediction for joint signal--idler discrimination}
\label{sec:joint_lda_theory}

The advantage of joint signal--idler discrimination is analogous to linear diversity combining in classical communication systems, where multiple noisy observations carrying the same underlying information are combined with suitable weights to maximize the output signal-to-noise ratio~\cite{Brennan1959Linear}. Here, the parametric interaction distributes the qubit-state information between the signal and idler, while also introducing correlations between their fluctuations. The 4D LDA projection plays the role of the combining rule: it weights the measured signal and idler quadratures according to their state-dependent separation and covariance to maximize the discrimination signal-to-noise separation.

To better understand the advantage of joint signal--idler discrimination, we model the TWPA as an ideal phase-preserving two-mode squeezer with power gain $G\geq1$. Internal TWPA losses in the signal and idler channels are represented by independent beam splitters with power transmissivities $\eta_s$ and $\eta_i$, respectively, placed after the ideal squeezing operation, following the lumped-loss model of Ref.~\cite{esposito2022observation}. The complete derivation is given in Appendix~\ref{app:lda_twpa_model}.

We consider a state-dependent coherent displacement at the signal input and vacuum at the idler input. Choosing local signal and idler quadratures so that the state-dependent displacement lies along the $I$ axes gives
\begin{equation}
    \Delta\mu = A
    \begin{pmatrix}
        \sqrt{\eta_s G}\\
        0\\
        \sqrt{\eta_i(G-1)}\\
        0
    \end{pmatrix},
    \label{eq:iib_mean_difference}
\end{equation}
where $A=|\Delta\alpha|$ is the distance between the two input signal
centroids in the quadrature convention used below. The common covariance
is
\begin{equation}
    \Sigma = \frac{1}{2}
    \begin{pmatrix}
        V_s & 0 & C & 0\\
        0 & V_s & 0 & -C\\
        C & 0 & V_i & 0\\
        0 & -C & 0 & V_i
    \end{pmatrix},
    \label{eq:iib_covariance}
\end{equation}
with
\begin{align}
    V_s &= N_{\mathrm{sys},s}+\frac{1}{2}+\eta_s(G-1), \nonumber\\
    V_i &= N_{\mathrm{sys},i}+\frac{1}{2}+\eta_i(G-1), \nonumber\\
    C &= \sqrt{\eta_s\eta_iG(G-1)}.
    \label{eq:iib_variance_definitions}
\end{align}
Here $N_{\mathrm{sys},s}$ and $N_{\mathrm{sys},i}$ are effective downstream added-noise contributions referred to the TWPA-output plane. Within the model, $C$ describes the signal--idler correlations generated by the parametric interaction, while the downstream noise is taken to be independent between the two frequency channels.

Applying LDA, the single-mode separations are
\begin{equation}
    R_s = \frac{2A^2\eta_sG}{V_s},
    \qquad
    R_i = \frac{2A^2\eta_i(G-1)}{V_i},
    \label{eq:iib_R2D}
\end{equation}
and the joint separation is
\begin{equation}
    R_{4\mathrm{D}} = \frac{2A^2}{V_sV_i-C^2} \left[ V_i\eta_sG+V_s\eta_i(G-1)-2C^2 \right].
    \label{eq:iib_R4D}
\end{equation}
The origin of the improvement becomes particularly transparent by rewriting Eq.~\eqref{eq:iib_R4D} as
\begin{equation}
    R_{4\mathrm{D}} = R_s + \frac{2A^2\left[d_i-(C/V_s)d_s\right]^2} {V_i-C^2/V_s},
    \label{eq:iib_R_conditional}
\end{equation}
where $d_s=\sqrt{\eta_sG}$ and $d_i=\sqrt{\eta_i(G-1)}$. In this expression, the factor $C/V_s$ quantifies how strongly fluctuations in the idler can be predicted from the signal. Consequently, $d_i-(C/V_s)d_s$ represents the component of the qubit-dependent idler displacement that is not redundant with the signal measurement, while $V_i-C^2/V_s$ is the corresponding residual idler-noise variance after removing the part linearly correlated with the signal. The second term in Eq.~\eqref{eq:iib_R_conditional} therefore quantifies the additional state-discrimination information carried by the idler after accounting for what is already contained in the signal and is necessarily non-negative. In this sense, joint readout is useful whenever the idler retains qubit-state information that is not fully predictable from the noisy signal record; independent downstream noise in the two frequency channels can contribute to such nonredundancy, while the qubit-dependent displacement is generated coherently in both modes.

\section{Experimental Analysis}
\label{sec:sec3_Experimental_Results}

We experimentally investigate the information carried by the signal and idler, the benefit of their joint discrimination, and the adaptation of the 4D linear discriminant to the TWPA pump phase. The measurements were performed on a superconducting qubit readout platform with a Josephson TWPA as the first-stage amplifier. The experimental setup, device, acquisition protocol, and train/test procedure are described in Appendix~\ref{app:device_and_setup}.

\subsection{Separate signal and idler readout versus integration time}
\label{sec:experimental_separate_readout}

\begin{figure*}[t]
    \centering
    \includegraphics[width=0.9\textwidth]
    {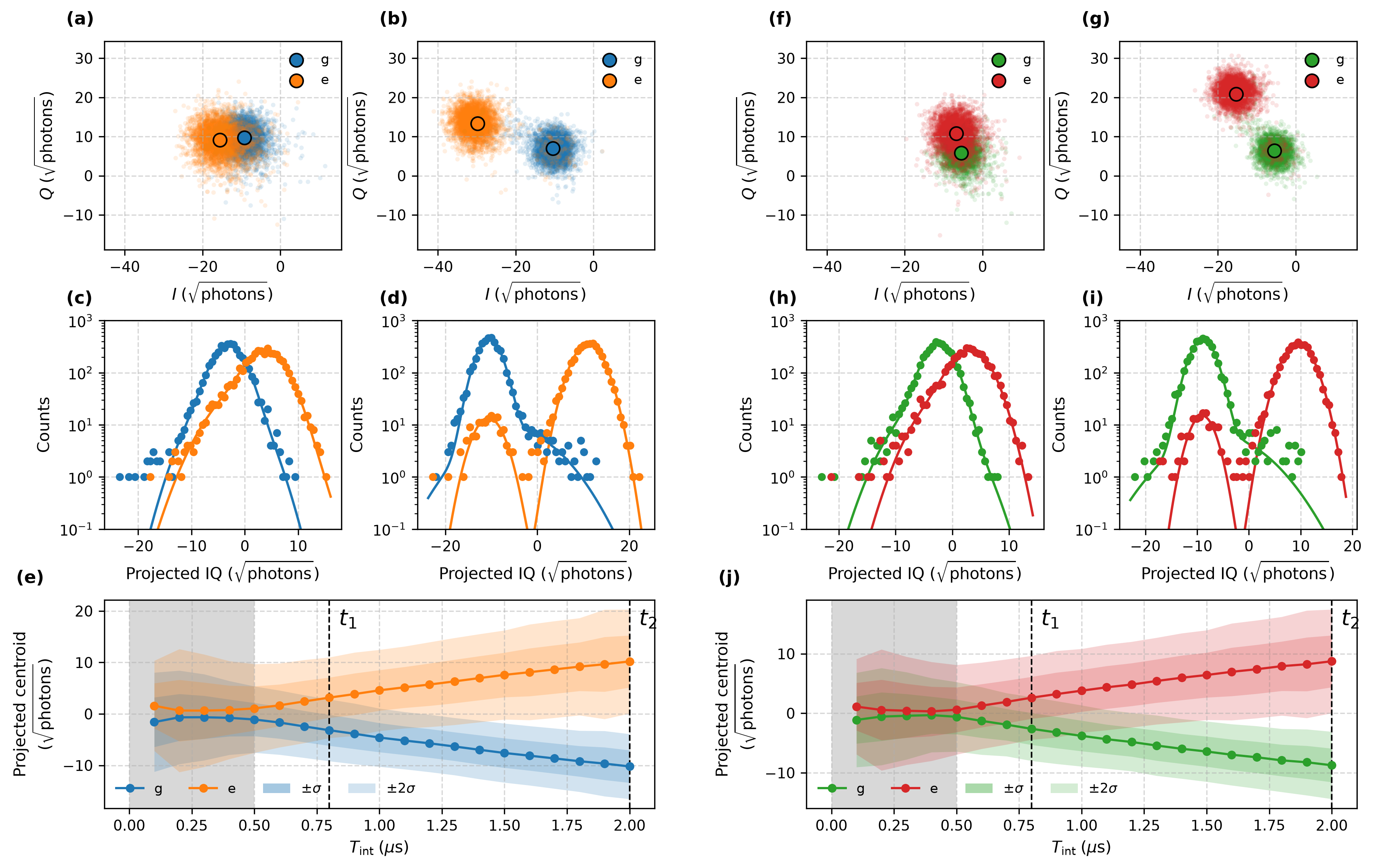}
    \caption{
    Signal and idler readout as a function of integration time. (a,b) Signal IQ distributions at $t_1=0.8\,\mu\mathrm{s}$ and $t_2=2.0\,\mu\mathrm{s}$, respectively; (f,g) corresponding idler distributions. Large markers indicate the state-conditioned centroids. (c,d) Histograms of the signal data in (a,b), projected onto the corresponding centroid-separation axis (dots), together with double-Gaussian fits (solid lines); (h,i) corresponding idler histograms. (e,j) Fitted projected centroids for the signal and idler, respectively, as a function of integration time, with shaded regions indicating the $\pm\sigma$ and $\pm2\sigma$ intervals obtained from single-Gaussian fits. Vertical dashed lines mark $t_1$ and $t_2$, and the gray shaded region indicates the initial noise-dominated acquisition window. IQ amplitudes are referred to the TWPA output plane.
    }
    \label{fig:signal_idler_iq_vs_integration_time}
\end{figure*}

We first establish that the idler carries a measurable qubit-state-dependent displacement. Figures~\ref{fig:signal_idler_iq_vs_integration_time}(a) and (b) show the signal IQ distributions at integration times $t_1=0.8\,\mu\mathrm{s}$ and $t_2=2.0\,\mu\mathrm{s}$; Figs.~\ref{fig:signal_idler_iq_vs_integration_time}(f) and (g) show the corresponding idler distributions calibrated to the TWPA's output port (see Appendix~\ref{app:quadratures_and_output_calibration}). Both modes exhibit separated ground- and excited-state clusters, with a larger integrated displacement at the longer acquisition time.

To quantify their separation, we project each distribution onto the axis joining its two state centroids. The resulting histograms and double-Gaussian fits are shown for the signal in
Figs.~\ref{fig:signal_idler_iq_vs_integration_time}(c) and (d) and for the idler in Figs.~\ref{fig:signal_idler_iq_vs_integration_time}(h) and (i). A small population extends into the opposite state's distribution, particularly for nominal excited-state preparation at longer integration times. Such departures from a single-Gaussian description are consistent with contributions from state-preparation errors or transitions during acquisition.

Figures~\ref{fig:signal_idler_iq_vs_integration_time}(e) and (j) track the fitted projected centroids as a function of integration time. The readout pulse and acquisition window are initiated simultaneously, without compensation for propagation delays in the measurement chain, internal acquisition latency, or the readout-resonator ring-up time. Consequently, the initial shaded interval contains little state-dependent dispersive signal and is dominated by measurement noise and transients. Beyond this interval, the separation between the ground- and excited-state distributions increases with integration time in both channels. At $T_{\mathrm{int}}=2\,\mu\mathrm{s}$, the signal- and idler-only classifiers achieve held-out assignment fidelities of $F_s=(97.35\pm0.29)\,\%$ and $F_i=(97.34\pm0.28)\,\%$, respectively. These results demonstrate that the idler can be acquired simultaneously with the signal and, under these operating conditions, carries qubit-state information sufficient to achieve discrimination performance comparable to that of the signal.

\subsection{Joint signal--idler discrimination}
\label{sec:experimental_joint_discrimination}

\begin{figure}[t]
    \centering
    \includegraphics[width=\linewidth]
    {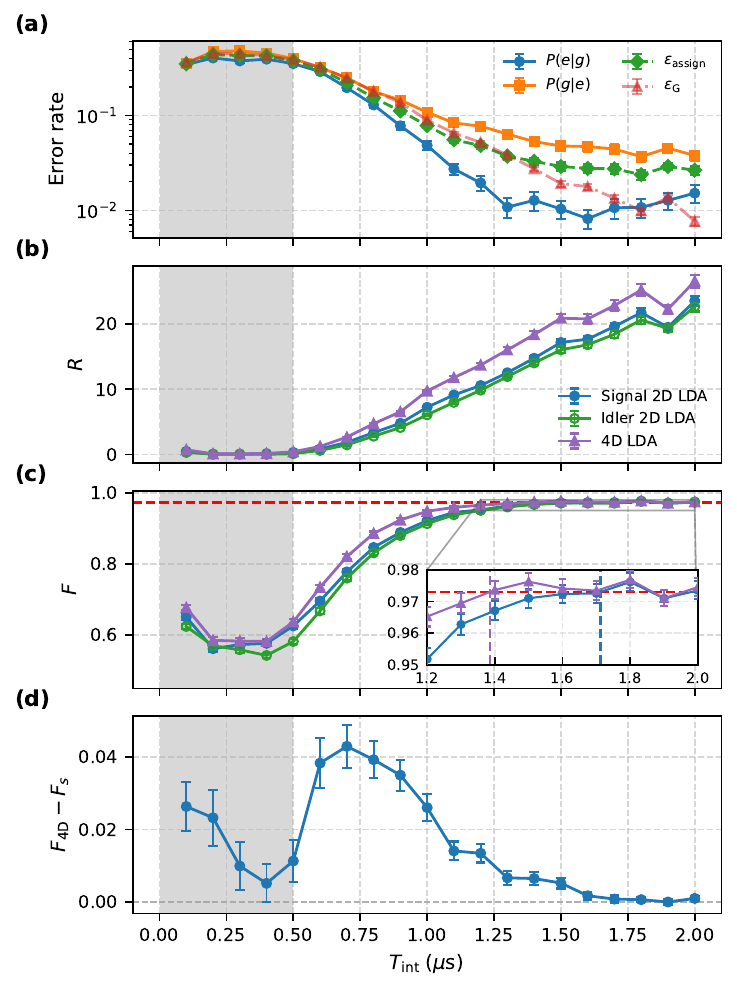}
    \caption{
    Discrimination performance as a function of integration time. (a) Signal-only assignment errors $P(e|g)$ and $P(g|e)$, balanced assignment error $\epsilon_{\mathrm{assign}}$, and Gaussian-overlap error $\epsilon_{\mathrm{G}}$. (b) LDA separation $R$ for signal-only, idler-only, and joint 4D discrimination; $R$ is calculated from the training-set LDA fit (c) Corresponding held-out assignment fidelities. The inset enlarges the high-fidelity region where the measured fidelities approach $F_{\mathrm{th}}=0.973$. The horizontal red dashed line marks $F_{\mathrm{th}}$, while the vertical dashed lines indicate the interpolated integration times at which the signal-only and 4D fidelity curves reach this value. (d) Paired fidelity difference $F_{\mathrm{4D}}-F_s$. The shaded region below $0.5\,\mu\mathrm{s}$ marks the initial noise- and transient-dominated acquisition window.
    }
    \label{fig:lda_vs_integration_time}
\end{figure}

We next compare signal-only, idler-only, and joint 4D LDA as a function of integration time. For each repeated train/test partition, $R$ is calculated from the LDA means and pooled covariance fitted on the training subset, while the assignment fidelity is evaluated on the held-out test shots. We first examine the signal-only assignment errors in Fig.~\ref{fig:lda_vs_integration_time}(a). The measured state-resolved errors are $P(e|g)$ and $P(g|e)$, with balanced assignment error
\begin{equation}
    \epsilon_{\mathrm{assign}} = \frac{P(e|g)+P(g|e)}{2}.
    \label{eq:experimental_assignment_error}
\end{equation}
For comparison, the Gaussian-overlap error inferred from the fitted separation is
\begin{equation}
    \epsilon_{\mathrm{G}} = \frac{1}{2} \operatorname{erfc}\!\left(\sqrt{\frac{R}{8}}\right).
    \label{eq:experimental_gaussian_error}
\end{equation}
Whereas $\epsilon_{\mathrm{G}}$ continues to decrease with increasing integration time, the measured assignment errors approach a plateau beyond approximately $1.3\,\mu\mathrm{s}$, showing that Gaussian cluster overlap alone does not account for the observed assignment errors in this regime.

Figure~\ref{fig:lda_vs_integration_time}(b) shows the discrimination separation $R$ for the three feature sets. The signal-only and idler-only classifiers use their respective two IQ quadratures, whereas the 4D classifier uses both modes simultaneously. The separation increases with integration time in all three cases, and the signal-only separation exceeds the idler-only separation throughout the measured range. This difference can reflect unequal channel transmission and added noise, as described by the analytical model in Sec.~\ref{sec:joint_lda_theory}. Joint 4D discrimination provides a larger separation than either individual channel, consistent with additional discrimination information in the joint signal--idler record.

The corresponding held-out assignment fidelities, $F=1-\epsilon_{\mathrm{assign}}$, are shown in Fig.~\ref{fig:lda_vs_integration_time}(c). At large integration times, the measured fidelities approach a plateau even though $R$ continues to increase. We define the plateau fidelity threshold $F_{\mathrm{th}}=0.973$ as the mean signal-only fidelity over $T_{\mathrm{int}}=1.6$--$2.0\,\mu\mathrm{s}$. From piecewise-linear interpolation of the mean held-out fidelity curves, the 4D classifier reaches this threshold at $T_{\mathrm{int}}\approx1.39\,\mu\mathrm{s}$, compared with $T_{\mathrm{int}}\approx1.71\,\mu\mathrm{s}$ for signal-only discrimination, corresponding to a reduction of approximately $326\,\mathrm{ns}$, or $19.1\,\%$. This threshold is shown in Fig.~\ref{fig:lda_vs_integration_time}(c). For comparison, Table~\ref{tab:integration_time_thresholds} gives the interpolated integration times for several additional fidelity thresholds. Across all thresholds considered, joint 4D discrimination reduces the required integration time by more than $10\,\%$, with the largest reduction occurring for the plateau threshold $F_{\mathrm{th}}=0.973$.

\begin{table}[t]
\centering
\caption{
Integration times required to reach selected fidelity thresholds $F_{\mathrm{th}}$, obtained by piecewise-linear interpolation of the mean held-out fidelity curves. Here $\Delta T=T_{\mathrm{int},s}-T_{\mathrm{int},4D}$, and the reduction is $1-T_{\mathrm{int},4D}/T_{\mathrm{int},s}$.
}
\label{tab}
\setlength{\tabcolsep}{5pt}
\begin{tabular}{ccccc}
\hline\hline
$F_{\mathrm{th}}$
& $T_{\mathrm{int},s}$ ($\mu$s)
& $T_{\mathrm{int},4D}$ ($\mu$s)
& $\Delta T$ (ns)
& Reduction (\%)\\
\hline
0.900 & 0.935 & 0.838 & 96  & 10.3 \\
0.950 & 1.175 & 1.020 & 156 & 13.2 \\
0.960 & 1.275 & 1.119 & 156 & 12.2 \\
0.970 & 1.474 & 1.313 & 161 & 10.9 \\
0.973 & 1.712 & 1.385 & 326 & 19.1 \\
\hline\hline
\end{tabular}
\label{tab:integration_time_thresholds}
\end{table}

Figure~\ref{fig:lda_vs_integration_time}(d) shows the paired fidelity difference $F_{\mathrm{4D}}-F_s$. The largest absolute fidelity improvement, reaching up to approximately four percentage points, occurs at shorter integration times where the overall fidelity is lower. As the integration time increases, this advantage gradually decreases and becomes negligible once the measured fidelity approaches its saturation plateau.

\subsection{Automatic adaptation to the pump phase}
\label{sec:experimental_pump_phase}

\begin{figure}[t]
    \centering
    \includegraphics[width=\linewidth]{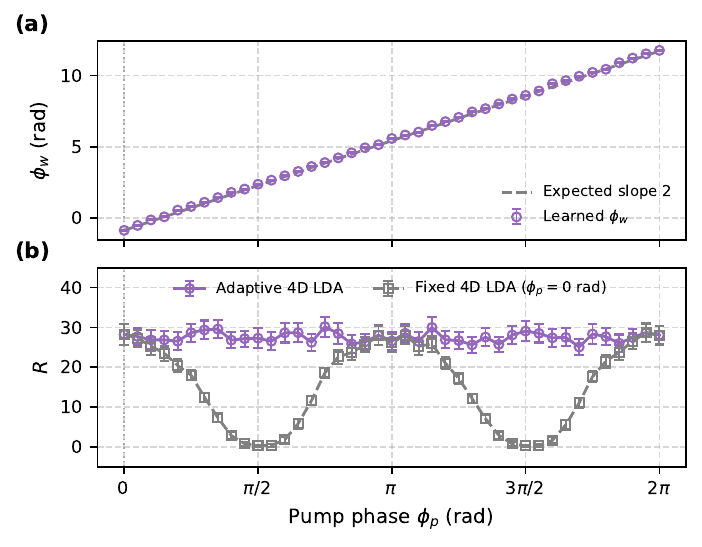}
    \caption{
    Joint-classifier adaptation versus TWPA pump phase. (a) Learned relative LDA weight phase $\phi_w$; the dashed line has slope $2$ and is anchored at the measured reference phase $\phi_p=0$. (b) Held-out projected separation $R_w$ for 4D LDA retrained at each pump phase and for a fixed classifier trained at $\phi_p=0$.}
    \label{fig:lda_pump_phase}
\end{figure}

Finally, we test whether 4D LDA adapts to controlled changes in the signal--idler phase relation without imposing a fixed quadrature-combination phase. We vary the TWPA pump phase $\phi_p$ and independently train a joint classifier at each setting. The four real LDA coefficients are represented compactly by
\begin{equation}
    w_s=w_{s,I}+i w_{s,Q},
    \qquad
    w_i=w_{i,I}+i w_{i,Q},
    \label{eq:experimental_complex_lda_weights}
\end{equation}
where these complex quantities are only a representation of pairs of real LDA weights; the classifier itself is trained on four real features. We characterize their relative orientation by
\begin{equation}
    \phi_w=\arg(w_i)-\arg(w_s).
    \label{eq:experimental_relative_weight_phase}
\end{equation}
For four-wave mixing, the parametric coupling phase satisfies $\varphi=2\phi_p+\varphi_0$, so the idler orientation changes twice as fast as the pump phase for a fixed signal phase. Accordingly, the expected relative weight phase is
\begin{equation}
    \phi_w(\phi_p)=\phi_w(0)+2\phi_p.
    \label{eq:experimental_expected_weight_phase}
\end{equation}
Figure~\ref{fig:lda_pump_phase}(a) shows that the learned phase follows this dependence. The plotted $\phi_w$ is unwrapped across the phase sweep, and the dashed reference is anchored to the learned phase at $\phi_p=0$ with slope fixed to $2$ by the four-wave-mixing phase relation rather than fitted to these data.

The corresponding behavior of the discrimination metric follows from the same rotation. To a good approximation, changing the pump phase rotates the idler IQ subspace by $\psi=2\phi_p$, so that
\begin{equation}
    \Delta\mu(\phi_p)=T(\psi)\Delta\mu(0),
    \qquad
    \Sigma(\phi_p)=T(\psi)\Sigma(0)T^T(\psi),
    \label{eq:pump_phase_rotation}
\end{equation}
where $T(\psi)$ acts as a rotation on the idler quadratures and leaves the signal quadratures unchanged. A retrained LDA direction transforms accordingly, $w(\phi_p)=T(\psi)w(0)$, leaving the projected separation invariant under this ideal rotation. By contrast, a classifier held fixed at $w_0$ gives
\begin{equation}
    R_{w_0}(\phi_p) =\frac{\left[w_0^T T(\psi)\Delta\mu(0)\right]^2}{w_0^T T(\psi)\Sigma(0)T^T(\psi)w_0},
    \label{eq:fixed_classifier_pump_phase}
\end{equation}
which varies periodically with $\phi_p$ as the idler rotates relative to the fixed classifier.

To test this experimentally, we compare the independently retrained classifier with a fixed classifier trained at $\phi_p=0$ and applied unchanged at every other pump phase. Both are evaluated on held-out shots using the projected separation
\begin{equation}
    R_w = \frac{(w^T\Delta\mu_{\mathrm{test}})^2}{w^T\Sigma_{\mathrm{test}}w}.
    \label{eq:experimental_projected_R}
\end{equation}
As shown in Fig.~\ref{fig:lda_pump_phase}(b), retraining maintains a nearly constant projected separation as the learned weights follow the rotating idler, whereas the fixed classifier exhibits the expected oscillatory degradation as it moves in and out of alignment with the measurement record. This demonstrates that LDA can adapt its signal and idler quadrature weights to controlled changes in the signal--idler phase relation, while possible contributions from pump-phase noise are discussed separately in Appendix~\ref{app:pump_phase_noise}.

\section{Discussion}
\label{sec:discussion}

Our measurements establish that the idler generated by a Josephson TWPA contains qubit-state information that can be exploited jointly with the signal to accelerate readout discrimination. When state-preparation, relaxation, and other non-Gaussian assignment errors are small, the single-shot measurement record is well described by the common-covariance Gaussian model, for which the joint separation satisfies $R_{4\mathrm{D}}\geq R_s$. Accordingly, when readout fidelity is limited primarily by overlap of the measurement distributions, 4D LDA is expected to reach a given target fidelity in a shorter integration time than signal-only discrimination.

This behavior is observed in our experiment, where joint discrimination reduces the required integration time by more than $10\,\%$ across the selected fidelity thresholds from $0.90$ to $0.973$. Additional assignment-error mechanisms limit the fidelity to $F_{\mathrm{th}}=0.973$, for which the reduction reaches approximately $19\,\%$. The absolute integration times, exceeding $1\,\mu\mathrm{s}$ near the maximum fidelity, are longer than current state-of-the-art values~\cite{Spring2025Fast}. This is primarily because neither the qubit architecture nor the readout power was optimized for fast readout. In addition, the readout pulse and acquisition window were initiated simultaneously, introducing a transient noise contribution that dominates at short integration times. These implementation-specific factors affect the absolute timescale but not the underlying benefit of retaining the idler, and we therefore expect the reduction in integration time to persist in faster, lower-error readout implementations.

The improvement obtained by retaining the idler is in line with the coherent signal--idler combination demonstrated in Ref.~\cite{Abdo2026enhancing}. Both approaches exploit the same signal--idler information but combine the two modes differently: the coherent method uses prescribed relative weights after phase alignment, whereas LDA learns the relative amplitude and phase weights from the measured centroids and covariance. In our measurement chain, the two methods give closely matched performance at the operating point used here, although a lower-noise chain can preserve a larger LDA advantage over the equal-amplitude coherent combination, as discussed in Appendix~\ref{app:lda_coherent_combination}. The main practical advantage of LDA in the present regime is therefore its data-driven implementation without requiring an explicit amplifier model or prescribed signal--idler combination.

Signal--idler discrimination requires simultaneous access to both output frequency bands and the corresponding acquisition and processing resources. It does not require any modification to the qubit or its readout interaction, but its implementation depends on whether the microwave and digitization chain can retain and process the idler. We demonstrate that this functionality can be implemented using currently available commercial qubit-control hardware. Likewise, 4D LDA can be implemented using standard Python packages, although it introduces two additional real-valued features relative to signal-only discrimination and requires estimation of a larger covariance matrix. Its computational cost and suitability for real-time feedback therefore depend on the acquisition hardware, training procedure, and specific implementation.

\section{Conclusion}
\label{sec:conclusion}

We have developed a loss-inclusive TWPA model within the LDA framework that shows how joint signal--idler measurements can provide additional discrimination information in standard dispersive qubit readout. We demonstrate experimentally that the generated idler carries qubit-state information and that combining the signal and idler quadratures with 4D LDA increases the discrimination separation. As a result, a given assignment-fidelity threshold can be reached with a shorter integration time than for signal-only discrimination. In our experiment, the reduction in integration time exceeds $10\,\%$ for all fidelity thresholds considered above $F=0.9$, reaching approximately $19\,\%$ at the observed plateau fidelity of $F=0.973$. We further show that the learned signal--idler weights adapt to controlled changes in the TWPA pump phase, illustrating the data-driven nature of the LDA approach: the relevant quadrature weights are determined directly from the measured state-dependent means and covariance rather than prescribed in advance. By exploiting information in the idler that would otherwise be discarded, this approach provides a practical route to reducing qubit-readout integration times using joint processing of the two TWPA output modes. Such reductions can be particularly valuable for repeated measurement and feedback operations in quantum error correction, where readout time contributes directly to the duration of the correction cycle.

\section*{Acknowledgments}
\label{sec:acknowledgments}

The authors thank Emilio Depero from Zurich Instruments for his support during the implementation of the measurement script.  This work was supported by the UK National Quantum Computing Centre [NQCC200921]. 


\appendix

\section{TWPA model with beam-splitter loss}
\label{app:lda_twpa_model}

\subsection{Amplifier transformation and state-dependent displacement}

We model the TWPA as an ideal phase-preserving nondegenerate parametric amplifier followed by independent effective loss channels for the signal and idler. The ideal two-mode transformation is \cite{caves1982quantum,weedbrook2012gaussian}
\begin{align}
    a_{s,0} &= \sqrt{G}\,a_{s,\mathrm{in}} + e^{i\varphi}\sqrt{G-1}\,a_{i,\mathrm{in}}^\dagger,\nonumber\\
    a_{i,0} &= \sqrt{G}\,a_{i,\mathrm{in}} + e^{i\varphi}\sqrt{G-1}\,a_{s,\mathrm{in}}^\dagger,
    \label{eq:lda_bogoliubov}
\end{align}
where $G\geq1$ is the ideal signal power gain, $\varphi$ is the phase of the parametric coupling, and the subscript $0$ denotes the output before the effective losses. For the four-wave-mixing process used here, the coupling phase follows $\varphi=2\phi_p+\varphi_0$, where $\phi_p$ is the applied pump phase and $\varphi_0$ is a fixed phase offset.

For each mode, we define
\begin{equation}
    I=\frac{a+a^\dagger}{2},
    \qquad
    Q=\frac{a-a^\dagger}{2i}.
    \label{eq:lda_quadrature_convention}
\end{equation}
A coherent state with amplitude $\alpha=\langle a\rangle$ therefore has mean quadratures $\langle I\rangle=\operatorname{Re}\alpha$ and $\langle Q\rangle=\operatorname{Im}\alpha$, while vacuum has variance $1/4$ in each quadrature.

\begin{figure}[t]
    \centering
    \includegraphics[width=\linewidth]{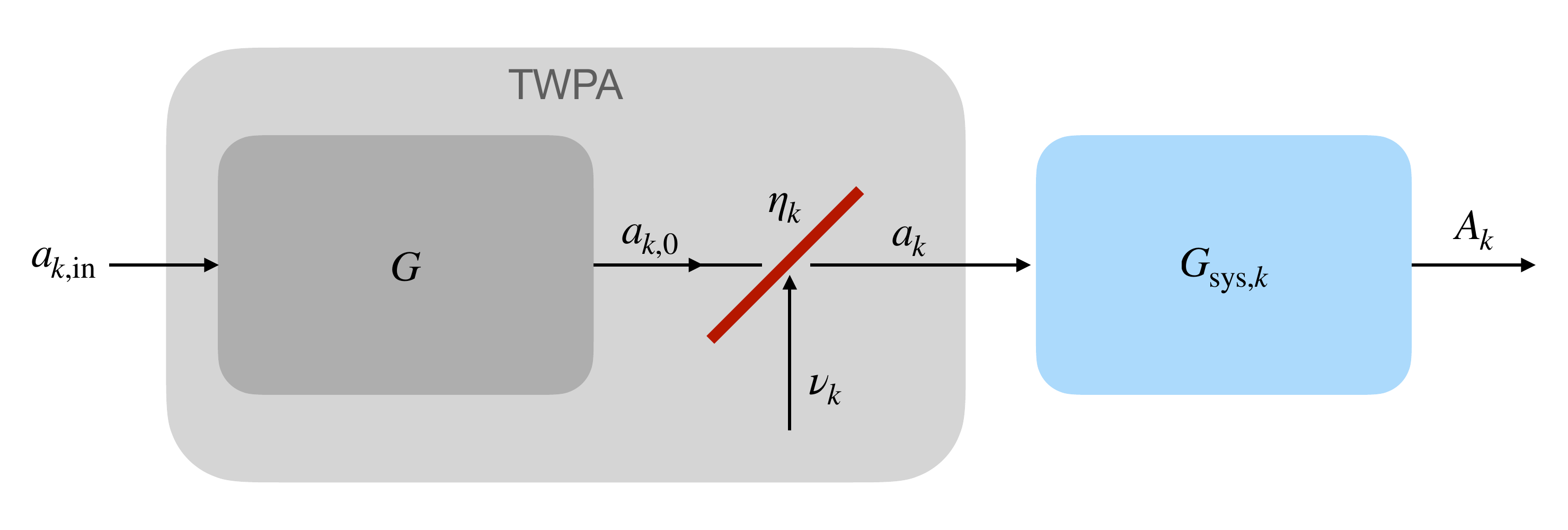}
    \caption{Schematic of the TWPA model with effective beam-splitter loss. Here $a_k$, with $k\in\{s,i\}$, denotes an electromagnetic field mode, $\eta_k$ is the corresponding beam-splitter power transmissivity, and $v_k$ is an independent vacuum mode.}
    \label{fig:lda_loss_ratio}
\end{figure}

Following the lumped-loss treatment of Ref.~\cite{esposito2022observation}, the effective signal and idler losses are represented by beam splitters placed after the ideal two-mode transformation,
\begin{align}
    a_s &= \sqrt{\eta_s}\,a_{s,0} + \sqrt{1-\eta_s}\,v_s, \nonumber\\
    a_i &= \sqrt{\eta_i}\,a_{i,0} + \sqrt{1-\eta_i}\,v_i,
    \label{eq:lda_splitter_loss}
\end{align}
where $v_s$ and $v_i$ are independent vacuum modes introduced by the loss channels. The parameters $\eta_s$ and $\eta_i$ are the corresponding beam-splitter power transmissivities, so the field amplitudes are attenuated by $\sqrt{\eta_s}$ and $\sqrt{\eta_i}$, respectively. The transformations in this model are illustrated in Fig.~\ref{fig:lda_loss_ratio}.

The qubit-state dependence is encoded in the coherent signal input. For prepared state $q\in\{g,e\}$, we define
\begin{equation}
    \alpha_q = \langle a_{s,\mathrm{in}}\rangle_q,
    \qquad
    \Delta\alpha = \alpha_e-\alpha_g,
    \qquad
    A=|\Delta\alpha|.
    \label{eq:lda_input_displacement}
\end{equation}
Thus, $A$ is the separation between the two input signal centroids, rather than the amplitude of either centroid individually. The idler input is assumed to have zero coherent amplitude for both qubit states, i.e., $\langle a_{i,\mathrm{in}}\rangle_q = 0$.

Taking the expectation value of Eq.~\eqref{eq:lda_bogoliubov} gives
\begin{align}
    \langle a_{s,0}\rangle_q &= \sqrt{G}\,\alpha_q, \nonumber\\
    \langle a_{i,0}\rangle_q &= e^{i\varphi}\sqrt{G-1}\,\alpha_q^*,
    \label{eq:lda_output_means}
\end{align}
and hence
\begin{align}
    \Delta\alpha_{s,0} &= \sqrt{G}\,\Delta\alpha, \nonumber\\
    \Delta\alpha_{i,0} &= e^{i\varphi}\sqrt{G-1}\,\Delta\alpha^*.
    \label{eq:lda_output_displacements}
\end{align}
The parametric interaction therefore maps the state-dependent input displacement onto both the amplified signal and the generated idler, with the idler displacement related to the signal by the phase-conjugating transformation.

For the analytical calculation, we independently rotate the signal and idler IQ coordinates so that each state-dependent displacement lies along its local $I$ axis, choosing the idler direction to give a positive displacement. After the effective losses, the difference between the two four-dimensional state centroids is then
\begin{equation}
    \Delta\mu = A
    \begin{pmatrix}
        \sqrt{\eta_sG}\\
        0\\
        \sqrt{\eta_i(G-1)}\\
        0
    \end{pmatrix}.
    \label{eq:lda_twpa_displacement}
\end{equation}
These rotations are only a choice of coordinates. For any block-diagonal orthogonal rotation $O$ acting independently on the signal and idler quadratures, $\Delta\mu\rightarrow O\Delta\mu$ and $\Sigma\rightarrow O\Sigma O^T$, leaving $\Delta\mu^T\Sigma^{-1}\Delta\mu$ unchanged. Experimental LDA can therefore be applied directly to the unrotated data.

\subsection{Output covariance matrix}

We next calculate the covariance of the output fluctuations. Since a coherent state at the amplifier input is simply a displacement of the vacuum state and has the same fluctuation statistics, the covariance can be evaluated using vacuum fluctuations at both amplifier inputs. For either mode, Eq.~\eqref{eq:lda_bogoliubov} gives
\begin{equation}
    \operatorname{Var}(I_{k,0}) = \operatorname{Var}(Q_{k,0}) = \frac{1}{4}+\frac{G-1}{2},
    \qquad 
    k\in\{s,i\}.
    \label{eq:lda_ideal_local_variance}
\end{equation}
After the beam-splitter loss, the vacuum entering through the unused port restores the vacuum-noise floor, giving
\begin{equation}
    \operatorname{Var}(I_k) = \operatorname{Var}(Q_k) = \frac{1}{4} + \frac{\eta_k(G-1)}{2}.
    \label{eq:lda_lossy_local_variance}
\end{equation}
These transformations are the standard covariance transformations for Gaussian loss channels~\cite{weedbrook2012gaussian}.

We additionally include downstream added noise through $N_{\mathrm{sys},s}$ and $N_{\mathrm{sys},i}$, referred to the effective TWPA-output plane. In our normalization, $N_{\mathrm{sys},k}$ denotes the added contribution to the sum of the two quadrature variances. Assuming this noise is isotropic and independent between the signal and idler channels, we define
\begin{align}
    V_s &= N_{\mathrm{sys},s} + \frac{1}{2} + \eta_s(G-1), \nonumber\\
    V_i &= N_{\mathrm{sys},i} +\frac{1}{2} + \eta_i(G-1),
    \label{eq:lda_local_noise}
\end{align}
such that
\begin{equation}
    \operatorname{Var}(I_k) = \operatorname{Var}(Q_k) = \frac{V_k}{2}.
    \label{eq:lda_V_variance_relation}
\end{equation}

The two-mode transformation also generates correlations between the signal and idler fluctuations. In the aligned coordinates used above, the ideal pre-loss covariances obtained from Eq.~\eqref{eq:lda_bogoliubov} are
\begin{align}
    \operatorname{Cov}(I_{s,0},I_{i,0}) &= \frac{\sqrt{G(G-1)}}{2}, \nonumber\\
    \operatorname{Cov}(Q_{s,0},Q_{i,0}) &= -\frac{\sqrt{G(G-1)}}{2},
    \label{eq:lda_ideal_cross_covariance}
\end{align}
while the cross-quadrature covariances vanish. The opposite signs reflect the phase-conjugating character of the two-mode transformation.

Because the two beam-splitter vacuum modes are independent, they do not contribute to the signal--idler cross-covariance. The correlations are therefore attenuated by $\sqrt{\eta_s\eta_i}$. Defining
\begin{equation}
    C = \sqrt{\eta_s\eta_iG(G-1)},
    \label{eq:lda_cross_covariance}
\end{equation}
we obtain
\begin{equation}
    \operatorname{Cov}(I_s,I_i)=\frac{C}{2},
    \qquad
    \operatorname{Cov}(Q_s,Q_i)=-\frac{C}{2}.
    \label{eq:lda_lossy_cross_covariance}
\end{equation}
Independent downstream added noise modifies the local variances but not these cross-covariances. The complete covariance matrix in the ordering $x=(I_s,Q_s,I_i,Q_i)^T$ is therefore
\begin{equation}
    \Sigma = \frac{1}{2}
    \begin{pmatrix}
        V_s & 0   & C   & 0\\
        0   & V_s & 0   & -C\\
        C   & 0   & V_i & 0\\
        0   & -C  & 0   & V_i
    \end{pmatrix}.
    \label{eq:lda_twpa_covariance}
\end{equation}
This covariance contains both the local signal and idler noise and the two-mode correlations generated by the parametric interaction.

\subsection{Signal-, idler-, and joint discrimination}

We now evaluate the squared Mahalanobis separation defined in Eq.~\ref{eq:lda_mahalanobis}. For compactness, we define
\begin{equation}
    d_s=\sqrt{\eta_sG},
    \qquad
    d_i=\sqrt{\eta_i(G-1)}.
    \label{eq:lda_ds_di_definition}
\end{equation}

For signal-only discrimination, $\Delta\mu_s=A(d_s,0)^T$ and $\Sigma_s=(V_s/2)\mathbb{I}_2$, giving
\begin{equation}
    R_s(G) = \frac{2A^2\eta_sG}{V_s} = \frac{2A^2\eta_sG}{N_{\mathrm{sys},s} + \frac{1}{2} + \eta_s(G-1)}.
    \label{eq:lda_R_signal}
\end{equation}
Similarly, for the idler, $\Delta\mu_i=A(d_i,0)^T$ and $\Sigma_i=(V_i/2)\mathbb{I}_2$, so that
\begin{equation}
    R_i(G) = \frac{2A^2\eta_i(G-1)}{V_i} = \frac{2A^2\eta_i(G-1)}{N_{\mathrm{sys},i} + \frac{1}{2} + \eta_i(G-1)}.
    \label{eq:lda_R_idler}
\end{equation}
The signal and idler squared centroid separations therefore scale with $G$ and $G-1$, respectively, while their discrimination performance is also set by the corresponding output noise.

For joint discrimination, the state-dependent mean difference has no $Q$ component in the aligned basis. Reordering the measurement vector as $x'=(I_s,I_i,Q_s,Q_i)^T$ makes the covariance matrix block diagonal, with separate $(I_s,I_i)$ and $(Q_s,Q_i)$ sectors. Since the centroid separation lies entirely in the $I$ sector, only the $(I_s,I_i)$ block contributes to the Mahalanobis separation,
\begin{equation}
    \Sigma_I = \frac{1}{2}
    \begin{pmatrix}
        V_s & C\\
        C   & V_i
    \end{pmatrix},
    \qquad
    \Delta\mu_I = A
    \begin{pmatrix}
        d_s\\
        d_i
    \end{pmatrix}.
    \label{eq:lda_I_covariance}
\end{equation}
Direct inversion of this $2\times2$ block gives
\begin{align}
    R_{\mathrm{4D}} &= \frac{2A^2}{V_sV_i-C^2} \left( V_i d_s^2 + V_s d_i^2 - 2Cd_sd_i \right)\nonumber\\
    &= \frac{2A^2}{V_sV_i-C^2}\left[ V_i\eta_sG + V_s\eta_i(G-1) - 2C^2 \right],
    \label{eq:lda_R_4d_intermediate}
\end{align}
where we used $d_s d_i=C$. Equation~\eqref{eq:lda_R_4d_intermediate} is the joint signal--idler separation including unequal signal and idler losses and downstream added noise. Equations~\eqref{eq:lda_R_signal}, \eqref{eq:lda_R_idler}, and \eqref{eq:lda_R_4d_intermediate} reduce to the corresponding lossless results for $\eta_s=\eta_i=1$. In the additional ideal limit $N_{\mathrm{sys},s}=N_{\mathrm{sys},i}=0$, the joint result becomes $R_{\mathrm{4D}}=4A^2$, independent of gain, as expected for an ideal unitary two-mode transformation.

The additional discrimination information provided by the idler can be made explicit by comparing the joint and signal-only separations. Using Eqs.~\eqref{eq:lda_R_4d_intermediate} and \eqref{eq:lda_R_signal},
\begin{align}
    R_{\mathrm{4D}}-R_s &= 2A^2 \left[\frac{V_i d_s^2+V_s d_i^2-2Cd_sd_i}{V_sV_i-C^2}-\frac{d_s^2}{V_s}\right]\nonumber\\
    &=\frac{2A^2\left(V_s d_i-Cd_s\right)^2}{V_s\left(V_sV_i-C^2\right)}.
    \label{eq:lda_R_difference_square}
\end{align}
For a physical covariance matrix, $V_sV_i-C^2>0$, so the improvement is non-negative, i.e., $R_{\mathrm{4D}}-R_s \ge 0$. Using $V_sV_i-C^2 =V_s\left(V_i-C^2/V_s\right)$, this can equivalently be written as
\begin{equation}
    R_{\mathrm{4D}} = R_s + \frac{2A^2\left[d_i-(C/V_s)d_s\right]^2}{V_i-C^2/V_s}.
    \label{eq:lda_R_conditional}
\end{equation}
This form shows that the joint separation consists of the signal-only contribution plus an additional non-negative term associated with the idler information not already correlated with the signal, as discussed in Sec.~\ref{sec:joint_lda_theory}. Within the model, retaining the idler therefore cannot reduce the optimal linear discrimination separation.

\section{Experimental setup and measurement procedure}
\label{app:device_and_setup}

The experiments were performed using a fixed-frequency superconducting transmon qubit on a chip designed and fabricated by Rigetti Computing, Inc~\cite{RigettiComputing}. The measured qubit and readout parameters, together with the operating frequencies, are summarized in Table~\ref{tab:qubit_parameters}. The first-stage amplifier was a commercially available Argo Josephson TWPA from Silent Waves~\cite{SilentWavesArgo}, operated in the four-wave-mixing regime.

\begin{table}[h]
    \caption{\label{tab:qubit_parameters}Qubit parameters and microwave frequencies used in the experiment.}
    \begin{ruledtabular}
    \begin{tabular}{lr}
        Parameter & Value \\
        \colrule
        \multicolumn{2}{l}{\textit{Qubit parameters}} \\
        Qubit frequency, $\omega_q/2\pi$ (MHz) & 3864.0 \\
        Anharmonicity magnitude, $|\alpha|/2\pi$ (MHz) & 114.3 \\
        Resonator linewidth, $\kappa/2\pi$ (MHz) & 1.8 \\
        Dispersive shift, $\chi/2\pi$ (MHz) & 1.1 \\
        Energy-relaxation time, $T_1$ ($\mu$s) & 74.1 \\
        \hline
        \multicolumn{2}{l}{\textit{TWPA parameters}} \\
        Signal frequency, $\omega_s/2\pi$ (MHz) & 5412.4 \\
        Idler frequency, $\omega_i/2\pi$ (MHz) & 6787.6 \\
        Pump frequency, $\omega_p/2\pi$ (MHz) & 6100.0 \\
    \end{tabular}
    \end{ruledtabular}
\end{table}

\begin{figure}[t]
    \centering
    \includegraphics[width=\linewidth]{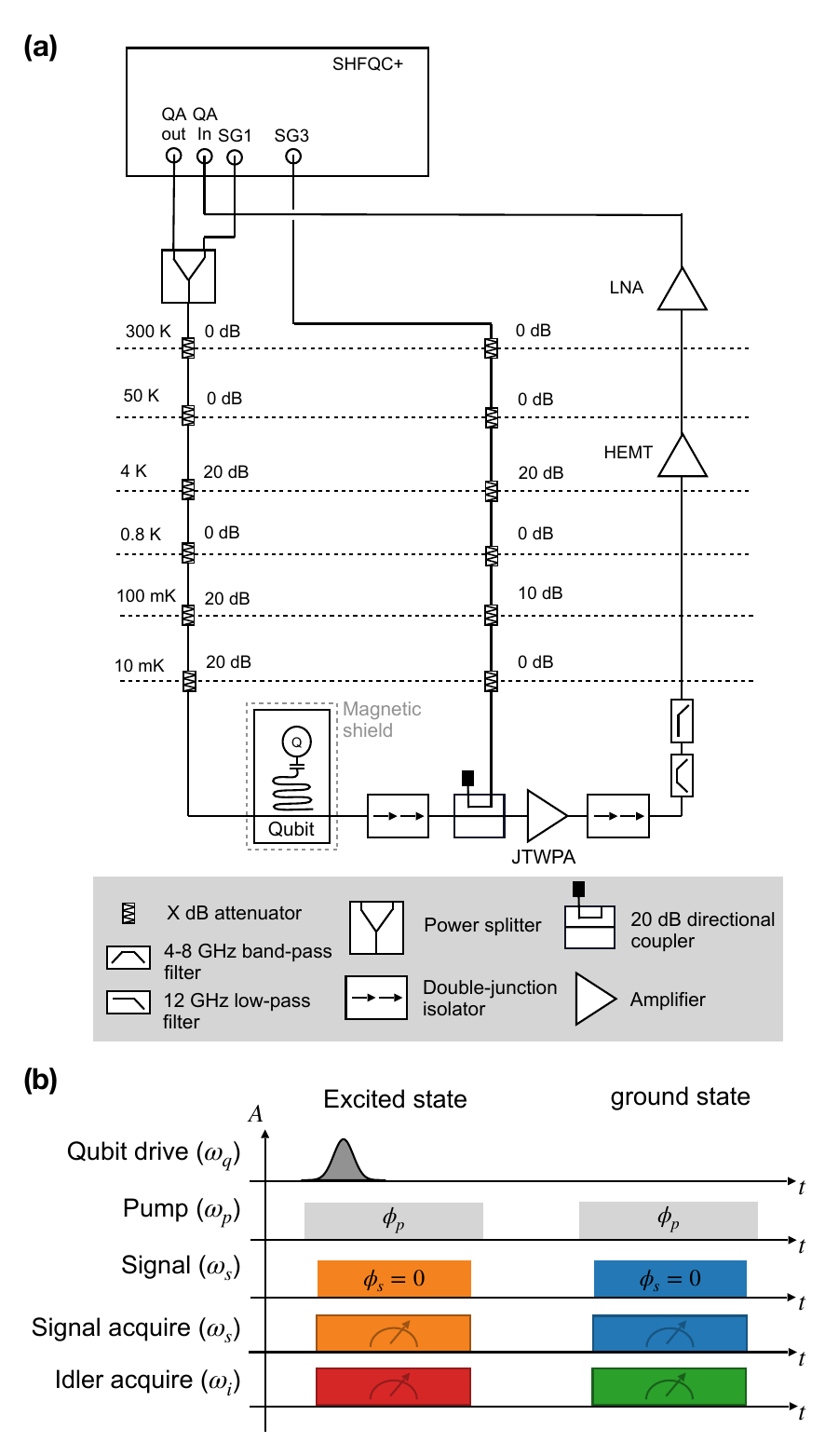}
    \caption{\label{fig:experimental_setup} (a) Experimental microwave readout chain. (b) Single-shot pulse sequence, including qubit control, TWPA pumping, readout, and simultaneous signal--idler acquisition. The pulse amplitudes and lengths are not to scale.}
\end{figure}

Figure~\ref{fig:experimental_setup}(a) shows the measurement chain. The qubit readout signal passes through a double-junction isolator (LNF-ISISC4\_8A) before being combined with the TWPA pump through a 20\,dB directional coupler (QMC-CRYOCOUPLER-20). Following the TWPA, a second double-junction isolator (LNF-ISISC4\_8A) suppresses noise propagating backwards from the HEMT amplifier (Low Noise Factory LNC4\_8C). The output chain also contains a 4--8\,GHz band-pass filter (Keenlion KBF-48-2S), a 12\,GHz low-pass filter (K\&L Microwave 6L250-00089), and a room-temperature low-noise amplifier (Pasternack PE15A1002). The qubit and TWPA were mounted on the mixing-chamber stage of a dilution refrigerator at approximately 10\,mK.

Microwave control and acquisition were performed using a Zurich Instruments SHFQC+~\cite{ZurichSHFQCPlus}, with experiments programmed through LabOne Q~\cite{ZurichLabOneQ}. The qubit-control and readout pulses were combined at room temperature and delivered through the common feedline. A separate signal-generator output channel supplied the pulsed TWPA pump and controlled its phase $\phi_p$ relative to the readout tone. Signal and idler were acquired simultaneously at $f_s=5412.4\,\mathrm{MHz}$ and $f_i=6787.6\,\mathrm{MHz}$, with $f_i=2f_p-f_s$.

The single-shot sequence is illustrated in Fig.~\ref{fig:experimental_setup}(b). The integration time $T_{\mathrm{int}}$ sets both the readout-pulse duration and acquisition window and was $2.0\,\mu\mathrm{s}$ unless otherwise stated. Each qubit-state acquired data set contained 4000 single-shot measurement records. The acquisition window begins at the same programmed time as the readout pulse, without compensating for the propagation delay or readout-resonator ring-up time; its initial portion therefore contains little dispersive signal and contributes to the noise- and transient-dominated region of the integration-time sweep.

For the discrimination results across the manuscript, we use 200 random 70/30 train/test partitions. Within each partition, the same state-resolved train/test indices are used for signal-only, idler-only, and joint classifiers. The LDA separation $R$ in the integration-time sweep is calculated from the class means and pooled covariance fitted on the training subset, whereas assignment fidelity is evaluated on the held-out test shots. Reported error bars are sample standard deviations across these repeated partitions and therefore quantify sensitivity to the data split; they are not uncertainties from independent experimental repetitions.

\section{Calibration of the measured field and output chain}
\label{app:quadratures_and_output_calibration}

\subsection{Effective transmission calibration}

The effective power transmissivities $\eta_s$ and $\eta_i$ used in the lumped-loss model are obtained from the TWPA insertion loss (IL) defined as the pump-off transmission relative to a nominally identical cryogenic reference line. Variation among three nominally identical reference lines is used to estimate the dominant calibration uncertainty. At the signal and idler frequencies,
\begin{equation}
    \mathrm{IL}_s=(3.75\pm1.98)\,\mathrm{dB},
    \qquad
    \mathrm{IL}_i=(4.76\pm1.96)\,\mathrm{dB},
    \label{eq:calibrated_splitter_transmissions}
\end{equation}
and
\begin{equation}
    \eta_k=10^{-\mathrm{IL}_k/10},
    \qquad
    \eta_s=0.4217,
    \qquad
    \eta_i=0.3342.
    \label{eq:eta_calibration}
\end{equation}
These should be interpreted as effective model transmissivities rather than direct measurements of intrinsic dissipative TWPA loss, because pump-off insertion loss can also contain mismatch, reflections, and other frequency-dependent effects. The use of separate cryogenic lines is an additional systematic limitation.

\begin{figure}
    \centering
    \includegraphics[width=\linewidth]{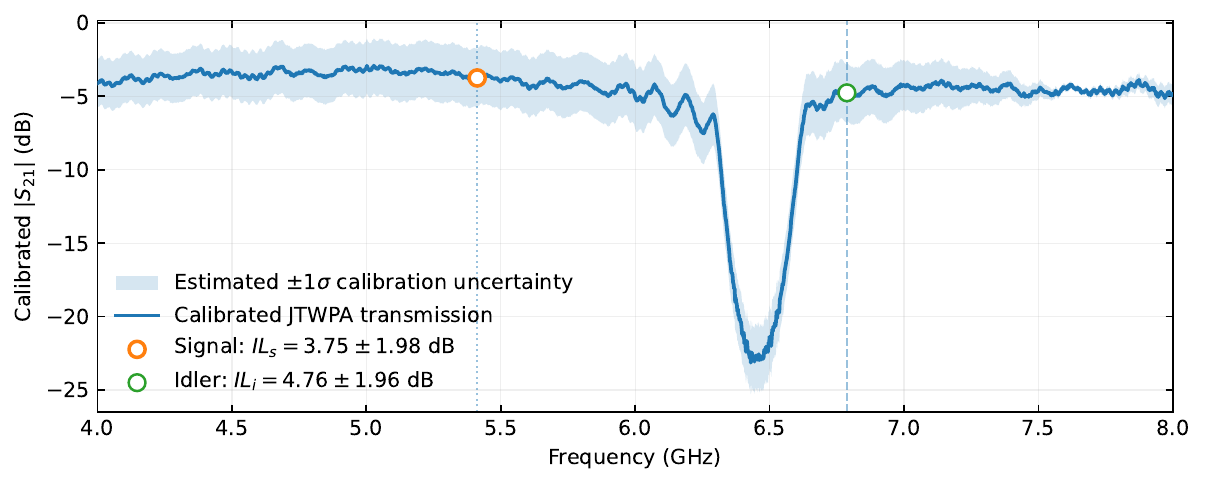}
    \caption{Calibrated pump-off insertion loss of the TWPA as a function of frequency. The shaded region indicates the estimated calibration uncertainty; markers identify the signal and idler frequencies.}
    \label{fig:twpa_insertion_loss}
\end{figure}

\subsection{Measured-quadrature calibration}
\label{app:quadrature_calibration}

Following Ref.~\cite{Abdo2026enhancing}, the SHFQC+ integration result $z_k^{(\mathrm{int})}$ is converted to output-referred quadratures in units of $\sqrt{\mathrm{photons}}$. For the constant integration kernel used here, the complete conversion can be written compactly as
\begin{equation}
    I_k+iQ_k = \frac{1}{\sqrt{G_{\mathrm{sys},k}}} \sqrt{\frac{T_{\mathrm{int}}}{Z_0\hbar\omega_k}} \frac{z_k^{(\mathrm{int})}}{r_{\mathrm{acq}}T_{\mathrm{int}}},
    \label{eq:output_referred_quadratures}
\end{equation}
where $Z_0=50\,\Omega$, $r_{\mathrm{acq}}=2\,\mathrm{GHz}$, and $G_{\mathrm{sys},k}$ is the independently calibrated downstream power gain. This removes the explicit integration-length dependence of the un-normalized SHFQC+ integration result before converting the measured field to the effective TWPA-output reference plane.

\subsection{Output-chain calibration}

We determine the effective downstream added noise $N_{\mathrm{sys},k}$ and gain $G_{\mathrm{sys},k}$, $k\in\{s,i\}$, from a sweep of the TWPA pump setting. The pump amplitude is controlled by the dimensionless SHFQC+ parameter $A_{\mathrm p}$, with the corresponding output range fixed at $10~\mathrm{dBm}$. Because the relation between $A_{\mathrm p}$ and the microwave power delivered to the TWPA is not independently calibrated, we use $A_{\mathrm p}$ itself as the experimental pump-control parameter.

For each value of $A_{\mathrm p}$, we acquire single-shot signal and idler measurements. The TWPA power gain is extracted from the coherent displacement of the ground-state signal distribution,
\begin{equation}
    G(A_{\mathrm p}) = \frac{\langle I_{s,g}(A_{\mathrm p})\rangle^2+\langle Q_{s,g}(A_{\mathrm p})\rangle^2}{\langle I_{s,g}(0)\rangle^2+\langle Q_{s,g}(0)\rangle^2}.
    \label{eq:measured_twpa_gain}
\end{equation}
We use the ground-state distribution because, at sufficiently long integration times, it is better described by the single-Gaussian model used in this calibration than the nominal excited-state distribution, consistent with the smaller measured $P(e|g)$ than $P(g|e)$ in Fig.~\ref{fig:lda_vs_integration_time}. Within the lumped beam-splitter loss model, the symmetrized noise referred to the lossy TWPA output is
\begin{equation}
    N_{\mathrm{TWPA},k}(G) = \frac{1}{2}+\eta_k(G-1).
    \label{eq:lossy_twpa_noise}
\end{equation}
A pump-independent signal transmission $\eta_s$ cancels from Eq.~\eqref{eq:measured_twpa_gain}, so the measured pump-on/pump-off coherent-power ratio still gives the parametric gain $G$ within this effective model.

\begin{figure}
\centering
\includegraphics[width=\linewidth]{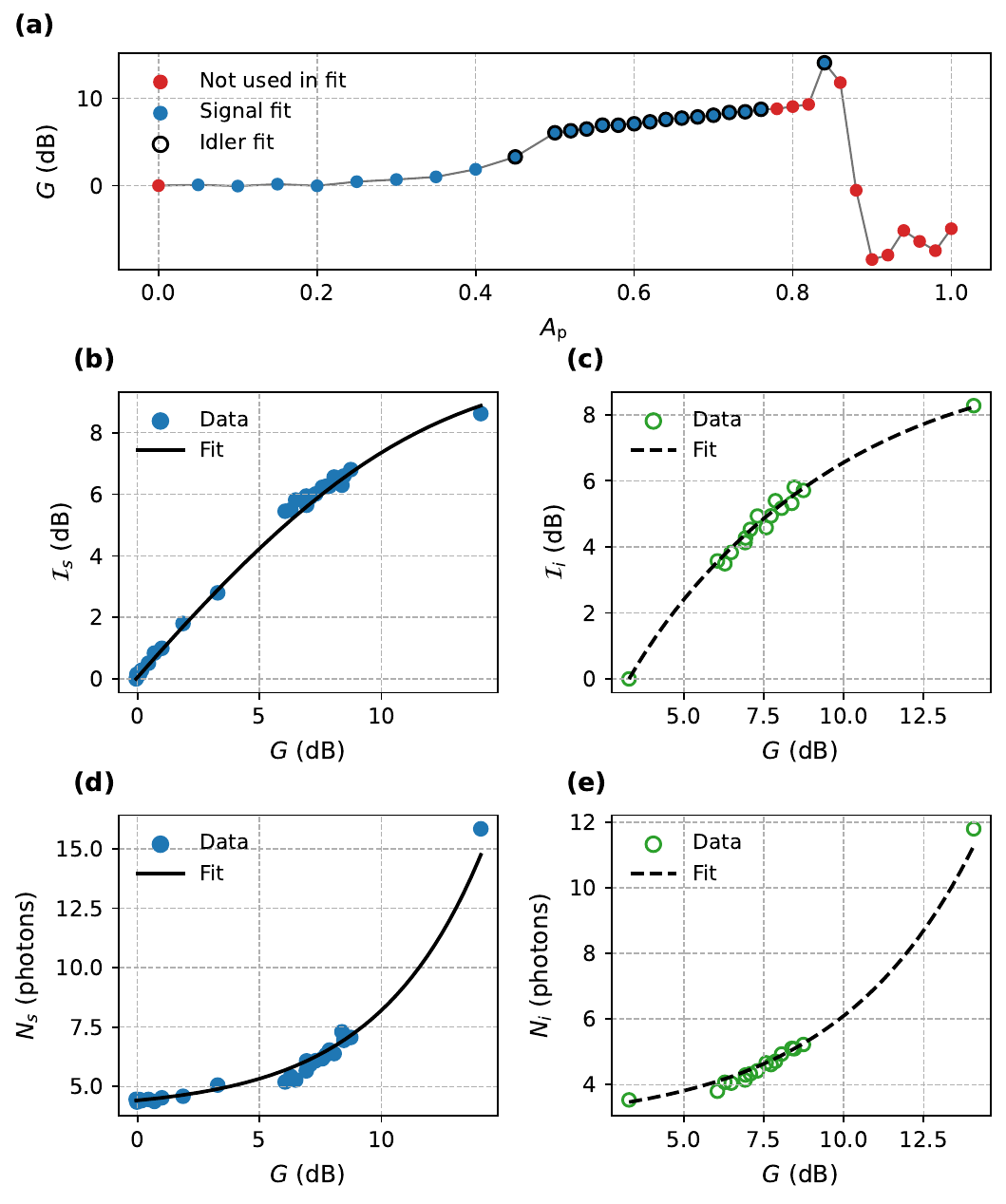}
\caption{Output-chain calibration. (a) Measured TWPA gain versus pump setting. (b,c) Normalized calibration metric $\mathcal{I}_k$ for the signal and idler with model fits. (d,e) Output-referred noise and fitted gain-dependent models.}
\label{fig:Output_Calibration}
\end{figure}

The complete measured gain sweep is shown in Fig.~\ref{fig:Output_Calibration}(a). Some pump settings are excluded from the calibration fits because phase instabilities produce single-shot IQ distributions that are poorly described by a single Gaussian, making the extracted means and variances unreliable for this procedure. We also exclude settings near or above the device critical-current regime, where dissipation becomes appreciable and the parametric gain decreases or disappears. The retained signal and idler points are indicated separately in Fig.~\ref{fig:Output_Calibration}(a).

To avoid confusion with the discrimination SNR defined in Eq.~\eqref{eq:signal_snr}, we characterize each single-Gaussian distribution using the calibration-specific coherent response-to-noise metric
\begin{equation}
    \mathcal{C}_k(G) = \frac{|\mu_k(G)|^2}{\sigma_{I,k}^2(G) + \sigma_{Q,k}^2(G)},
    \qquad
    |\mu_k|^2 = \langle I_k\rangle^2 + \langle Q_k\rangle^2,
    \label{eq:calibration_metric}
\end{equation}
and its normalized improvement relative to a reference point,
\begin{equation}
    \mathcal{I}_k(G) = \frac{\mathcal{C}_k(G)}{\mathcal{C}_k(G_{\mathrm{ref},k})}.
    \label{eq:calibration_improvement}
\end{equation}
The reference gain is chosen independently for the two channels. For the signal, whose coherent power scales as $G$, a point near unity gain is well conditioned. For the idler, however, the generated coherent power scales as $G-1$, so choosing $G_{\mathrm{ref},i}$ too close to unity would result in a very small reference displacement and amplify its uncertainty through the normalization. We therefore choose an idler reference gain appreciably larger than unity.

Using the coherent-power scalings $\eta_sG$ and $\eta_i(G-1)$ for the signal and idler, respectively, the normalized calibration models are
\begin{align}
    \mathcal{I}_s(G) &= \frac{G}{G_{\mathrm{ref},s}} \frac{N_{\mathrm{TWPA},s}(G_{\mathrm{ref},s})+N_{\mathrm{sys},s}}{N_{\mathrm{TWPA},s}(G)+N_{\mathrm{sys},s}},\nonumber\\
    \mathcal{I}_i(G) &= \frac{G-1}{G_{\mathrm{ref},i}-1} \frac{N_{\mathrm{TWPA},i}(G_{\mathrm{ref},i})+N_{\mathrm{sys},i}}{N_{\mathrm{TWPA},i}(G)+N_{\mathrm{sys},i}}.
    \label{eq:calibration_improvement_models}
\end{align}
The measured ground-state values of $\mathcal{I}_s$ and $\mathcal{I}_i$ are shown in Figs.~\ref{fig:Output_Calibration}(b) and (c). Fits to Eq.~\eqref{eq:calibration_improvement_models} yield $N_{\mathrm{sys},s}$ and $N_{\mathrm{sys},i}$.

The absolute downstream gain is then obtained from the measured noise,
\begin{equation}
    N_k^{(\mathrm{meas})}(G) = G_{\mathrm{sys},k} \left[ N_{\mathrm{TWPA},k}(G)+N_{\mathrm{sys},k} \right],
    \label{eq:absolute_noise_model}
\end{equation}
with $N_{\mathrm{sys},k}$ and $\eta_k$ fixed from the previous calibration steps. Once $G_{\mathrm{sys},k}$ is known, the measured noise is referred back to the effective TWPA-output plane as
\begin{equation}
    N_k(G) = \frac{N_k^{(\mathrm{meas})}(G)}{G_{\mathrm{sys},k}} = N_{\mathrm{TWPA},k}(G)+N_{\mathrm{sys},k},
    \label{eq:output_referred_noise}
\end{equation}
where experimentally $N_k=\sigma_{I,k}^2+\sigma_{Q,k}^2$ after applying the quadrature calibration of Eq.~\eqref{eq:output_referred_quadratures}. The resulting output-referred noise is shown in Figs.~\ref{fig:Output_Calibration}(d) and (e).

The statistical uncertainties from the calibration fits are smaller than the uncertainty associated with the TWPA insertion loss. We therefore repeat the full calibration using the upper and lower insertion-loss limits of Eq.~\eqref{eq:calibrated_splitter_transmissions} and treat the resulting variation as a systematic uncertainty. The fit and insertion-loss contributions are combined in quadrature to obtain the intervals reported in Table~\ref{tab:output_chain_calibration}.

For completeness, we express the equivalent system added-noise photon number as an effective noise temperature,
\begin{equation}
    T_{\mathrm{sys},k} = \frac{\hbar\omega_k}{k_{\mathrm B}} N_{\mathrm{sys},k}.
    \label{eq:system_noise_temperature}
\end{equation}
The central values are $T_{\mathrm{sys},s}=1.017~\mathrm{K}$ and $T_{\mathrm{sys},i}=0.843~\mathrm{K}$, with total uncertainty intervals of approximately $[0.596,1.681]~\mathrm{K}$ and $[0.477,1.418]~\mathrm{K}$, respectively. These values are of the same order as the approximately $1.5~\mathrm{K}$ added-noise temperature expected from the HEMT amplifier, which is anticipated to dominate the downstream noise. The lower central values may reflect the limited number of high-gain calibration points, uncertainty in the insertion-loss estimate, and the simplifying assumptions of the lumped-loss model. Accordingly, $N_{\mathrm{sys},k}$ and $T_{\mathrm{sys},k}$ should be interpreted as effective first-order estimates of the downstream noise rather than precise measurements of its individual physical contributions. This accuracy is sufficient for the present analysis, where these parameters are used to characterize the approximate noise levels of the two measurement channels.

\begin{table}[t]
    \caption{Output-chain calibration parameters. Uncertainties combine the fit and insertion-loss contributions in quadrature.}
    \label{tab:output_chain_calibration}
    \begin{ruledtabular}
        \begin{tabular}{lcccc}
            Mode
            & $G_{\mathrm{sys},k}$
            & $G_{\mathrm{sys},k}$ (dB)
            & $N_{\mathrm{sys},k}$
            & $T_{\mathrm{sys},k}$ (K)
            \\
            \colrule
            Signal
            & $\left(1.64^{+0.95}_{-0.60}\right)\times10^{7}$
            & $72.16 \pm 1.98$
            & $3.92^{+2.55}_{-1.62}$
            & $1.02^{+0.66}_{-0.42}$
            \\
            Idler
            & $\left(2.47^{+1.41}_{-0.90}\right)\times10^{7}$
            & $73.93 \pm 1.96$
            & $2.59^{+1.76}_{-1.12}$
            & $0.84^{+0.57}_{-0.37}$
            \\
        \end{tabular}
    \end{ruledtabular}
\end{table}

\section{Analysis of pump-phase fluctuations and anisotropic noise}
\label{app:pump_phase_noise}

\begin{figure}
    \centering
    \includegraphics[width=\linewidth]{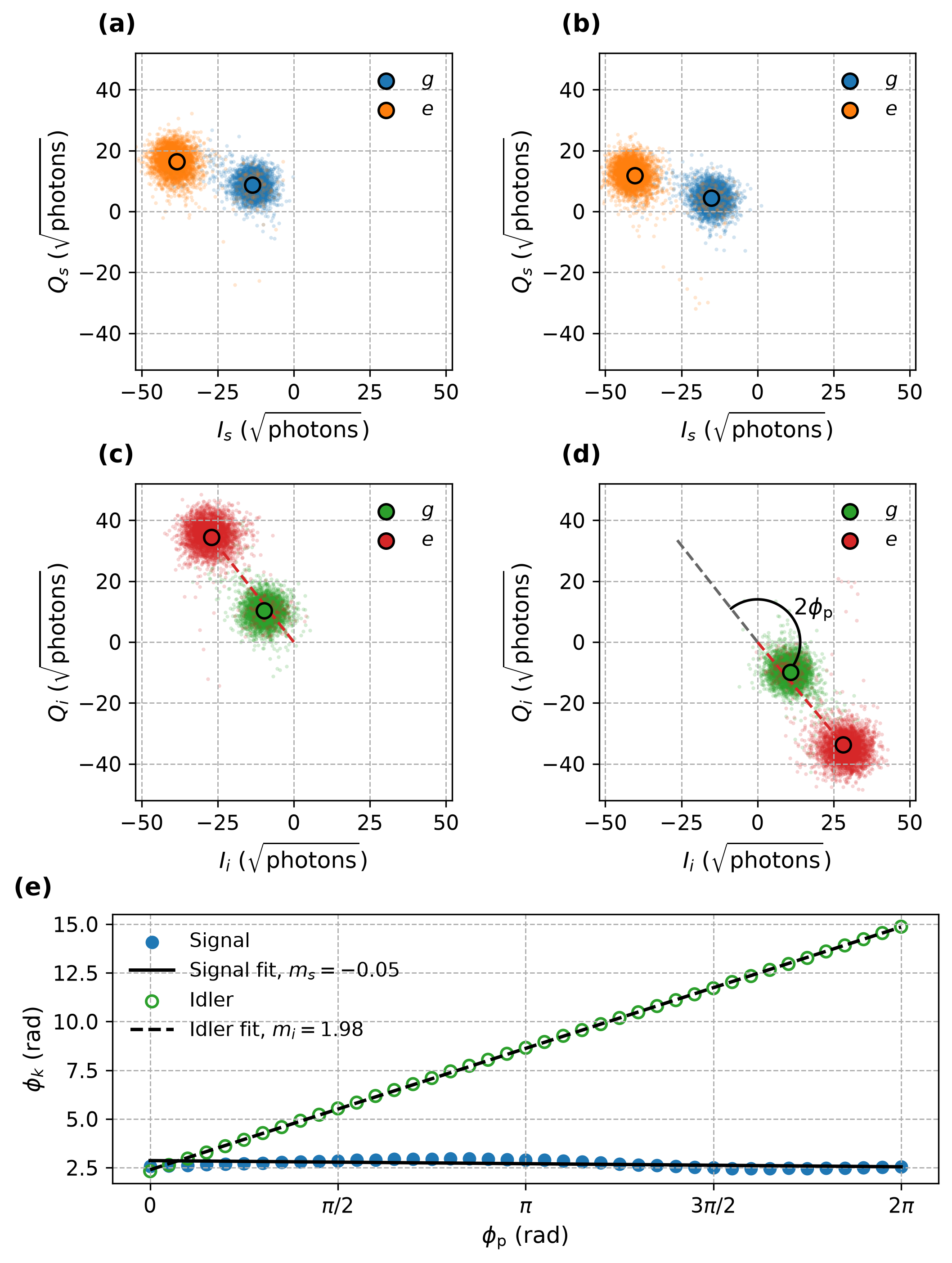}
    \caption{Dependence of the signal and idler fields on pump phase. (a,b) Signal IQ distributions for two pump phases separated by approximately $\pi/2$. (c,d) Corresponding idler IQ distributions. (e) Ground-state centroid phase versus pump phase; the idler fit has slope $m_i=1.98$, consistent with the four-wave-mixing expectation of $2$.}
    \label{fig:pump_phase_IQ_blobs}
\end{figure}

For four-wave mixing, the generated idler phase obeys
\begin{equation}
    \phi_i=2\phi_p-\phi_s+\phi_0,
    \label{eq:idler_phase_relation}
\end{equation}
so for fixed signal phase the idler rotates twice as fast as the pump. Figure~\ref{fig:pump_phase_IQ_blobs} shows this behavior directly. Panels (a,b) and (c,d) show the signal and idler IQ distributions, respectively, for two pump phases separated by approximately $\pi/2$. The signal orientation remains approximately unchanged, whereas the idler rotates by approximately $\pi$. As shown in panel (e), a linear fit to the ground-state centroid phase gives an idler slope $m_i=1.98$, consistent with the expected value of $2$, while the signal phase is approximately independent of $\phi_p$. This strong idler-phase dependence motivates a check of whether pump-phase fluctuations could contribute to anisotropic idler noise.

To investigate the possible contribution of pump-phase noise, we define the tangential excess variance for each mode as
\begin{equation}
D_{\theta,k} = \sigma_{\theta,k}^2-\sigma_{r,k}^2,
\qquad
k\in{s,i},
\label{eq:tangential_excess}
\end{equation}
where $\sigma_{r,k}^2$ and $\sigma_{\theta,k}^2$ are the variances parallel and perpendicular, respectively, to the coherent IQ displacement. This quantity provides a measure of noise anisotropy: isotropic noise contributes equally to both variances and therefore cancels in their difference, whereas phase fluctuations preferentially broaden the distribution in the tangential direction, contributing positively to $D_{\theta,k}$.

In the four-wave-mixing regime, a small pump-phase fluctuation $\delta\phi_p$ produces an idler-phase fluctuation $\delta\phi_i=2\delta\phi_p$. To leading order, the resulting tangential fluctuation is $|\alpha_i|\delta\phi_i$, giving a contribution to the idler tangential excess variance at the TWPA-output reference plane of
\begin{equation}
D_{\theta,i,\mathrm{pump}} \simeq 4|\alpha_i|^2\sigma_{\phi_p,\mathrm{eff}}^2,
\label{eq:idler_phase_noise_gain}
\end{equation}
where $\sigma_{\phi_p,\mathrm{eff}}^2$ is the effective pump-phase variance sampled during the measurement. Since $|\alpha_i|^2\propto\eta_i(G-1)$, the model predicts a positive contribution to the idler tangential excess variance that increases linearly with $G-1$, assuming a gain-independent effective pump-phase variance.

Experimentally, a direct covariance estimate of the ground-state cloud can be biased by a secondary population extending toward the excited-state distribution. We therefore fit the nominal ground-state data with a two-component Gaussian mixture along the ground--excited separation axis and use the resulting posterior ground-like probabilities as statistical weights. These weights are then applied to the original two-dimensional IQ data to calculate the weighted covariance matrix. The radial axis is defined along the ground-state coherent displacement and the tangential axis orthogonal to it, from which $\sigma_{r,k}^2$, $\sigma_{\theta,k}^2$, and $D_{\theta,k}$ are obtained.

\begin{figure}
    \centering
    \includegraphics[width=\linewidth]{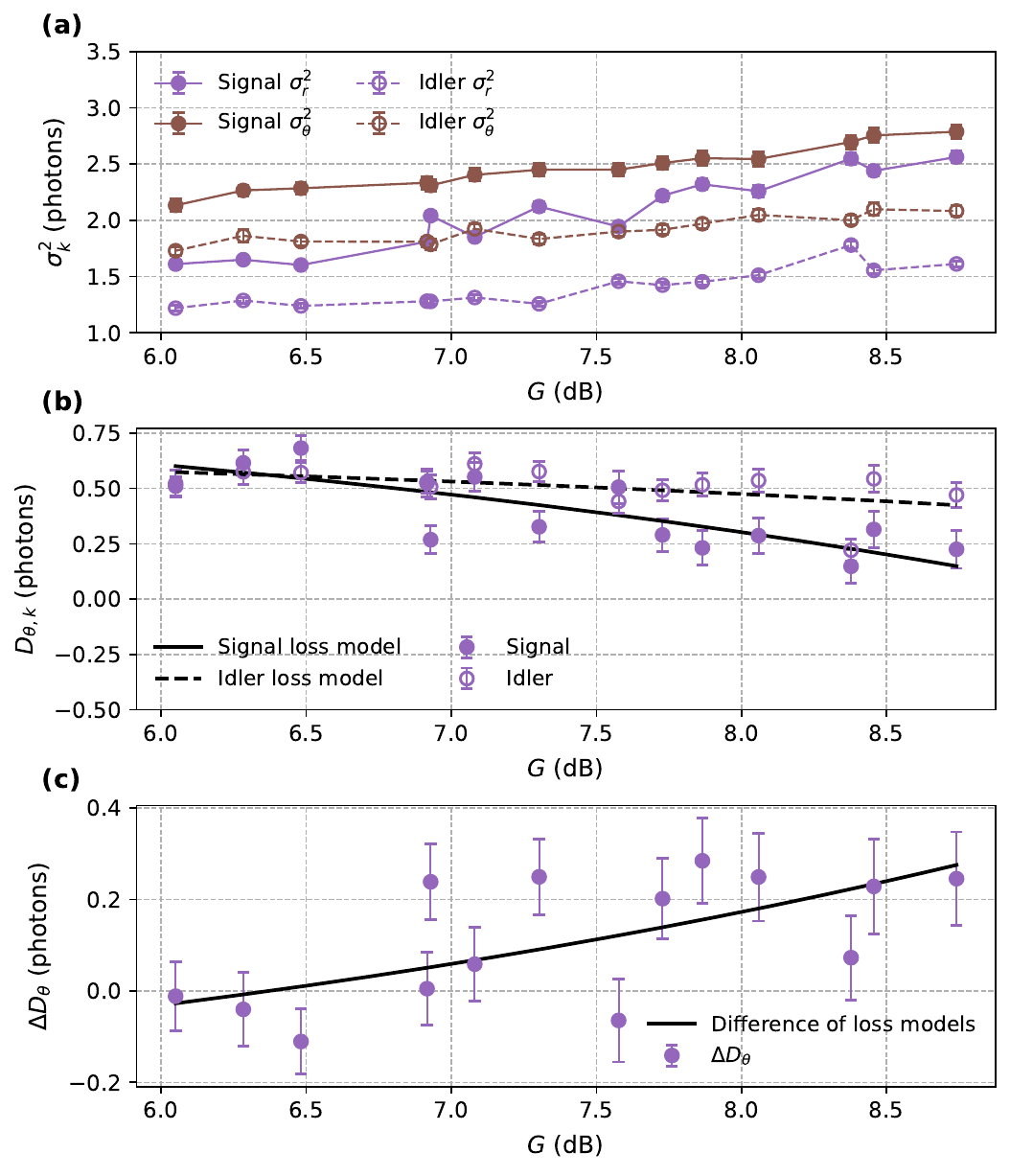}
    \caption{Radial and tangential quadrature-noise analysis. (a) Radial and tangential variances for both modes versus TWPA gain. (b) Tangential excess $D_{\theta,k}=\sigma_{\theta,k}^2-\sigma_{r,k}^2$ with the phenomenological fits of Eq.~\eqref{eq:tangential_splitter_fit}. (c) Differential excess $\Delta D_{\theta}=D_{\theta,i}-D_{\theta,s}$ and the difference between the two fitted curves. Error bars are bootstrap statistical uncertainties and do not include insertion-loss uncertainty.}
    \label{fig:pump_phase_anisotropy}
\end{figure}

The resulting radial and tangential variances are shown in Fig.~\ref{fig:pump_phase_anisotropy}(a). The tangential variance is generally larger than the radial variance for both modes, indicating that the observed anisotropy is not unique to the idler. Panel (b) shows the corresponding tangential excess, which we describe using the phenomenological model
\begin{equation}
    D_{\theta,k}(G) = D_{0,k}+\eta_k a_k(G-1),
    \label{eq:tangential_splitter_fit}
\end{equation}
with $\eta_k$ fixed by the insertion-loss calibration. Here $D_{0,k}$ is an offset, while $\eta_k a_k$ is the observable gain-dependent slope at the TWPA output reference plane. The fitted parameters are listed in Table~\ref{tab:anisotropy_fits}.

\begin{table}[t]
    \caption{Phenomenological anisotropy-fit parameters. The quoted uncertainties are statistical fit uncertainties conditional on the fixed calibration parameters.}
    \label{tab:anisotropy_fits}
    \begin{ruledtabular}
    \begin{tabular}{lccc}
        Mode & $D_{0,k}$ & $a_k$ & $\eta_k a_k$ \\
        \colrule
        Signal & $0.10\pm0.08$ & $-0.31\pm0.04$
               & $-0.13\pm0.02$ \\
        Idler  & $0.70\pm0.06$ & $-0.13\pm0.04$
               & $-0.04\pm0.01$ \\
    \end{tabular}
    \end{ruledtabular}
\end{table}

Figure~\ref{fig:pump_phase_anisotropy}(c) shows the differential excess
\begin{equation}
    \Delta D_{\theta} = D_{\theta,i}-D_{\theta,s}.
\end{equation}
The corresponding fitted slope difference is $\eta_i a_i-\eta_s a_s=0.09\pm0.02$ when the two fitted slope uncertainties are treated as independent. A limitation of this analysis is that the fraction of measurement points assigned to the ground-state population varies with gain, increasing from approximately $61\,\%$ to above $90\,\%$ for the signal and from approximately $59\,\%$ to about $90\,\%$ for the idler. Nevertheless, the positive differential slope indicates that the signal tangential excess variance decreases more rapidly with gain than that of the idler. However, both fitted slopes are negative, whereas the simple pump-phase-noise model predicts a positive idler contribution proportional to $G-1$. The measurements therefore reveal a small mode-dependent difference in noise anisotropy that cannot be unambiguously attributed to pump-phase noise.

\section{LDA and coherent signal--idler combination}
\label{app:lda_coherent_combination}

\subsection{Theoretical analysis}

We compare 4D LDA with the coherent signal--idler combination demonstrated in Ref.~\cite{Abdo2026enhancing}. The connection between the two methods is most transparent in the aligned coordinates of Appendix~\ref{app:lda_twpa_model}, where the state-dependent mean difference is $\Delta\mu=A(d_s,0,d_i,0)^T$.

The coherent-combination procedure of Ref.~\cite{Abdo2026enhancing} first rotates the measured signal and idler IQ distributions into aligned coordinates and refers both channels to common calibrated field units. A relative phase rotation $\phi$ is then applied to the idler record before the two fields are added quadrature by quadrature. Writing the rotated idler as
\begin{equation}
    I_i^{(r)} = \cos\phi\,I_i-\sin\phi\,Q_i,
    \qquad
    Q_i^{(r)} = \sin\phi\,I_i+\cos\phi\,Q_i,
\end{equation}
the collective information quadrature is
\begin{equation}
    y_{\mathrm{coh}}(\phi) = I_s+\cos\phi\,I_i-\sin\phi\,Q_i,
    \label{eq:coherent_combination_general}
\end{equation}
corresponding to
\begin{equation}
    w_{\mathrm{coh}}(\phi) =
    \begin{pmatrix}
        1\\
        0\\
        \cos\phi\\
        -\sin\phi
    \end{pmatrix}.
    \label{eq:lda_coh_phase_weight}
\end{equation}
The phase sweep therefore changes the orientation and sign of the idler contribution but not its magnitude, since $\cos^2\phi+\sin^2\phi=1$. After calibration to common field units, the prescribed coherent method optimizes the relative phase of an equal-amplitude signal--idler combination.

For the antisqueezed information quadrature of Ref.~\cite{Abdo2026enhancing}, the optimum phase aligns the state-dependent signal and idler displacements constructively. In the coordinate convention of Appendix~\ref{app:lda_twpa_model}, this phase alignment is already absorbed into the local idler axes, so that
\begin{equation}
    y_{\mathrm{coh}} = I_s+I_i,
    \qquad
    w_{\mathrm{coh}} =
    \begin{pmatrix}
        1\\
        0\\
        1\\
        0
    \end{pmatrix}.
    \label{eq:lda_coh_weight}
\end{equation}
The equal coefficients apply only after the two channels have been referred to common calibrated units and phase aligned; they do not represent equal weights on the raw measurement records.

Using the covariance parameters $V_s$, $V_i$, and $C$ defined in Appendix~\ref{app:lda_twpa_model}, the corresponding Gaussian separation is
\begin{equation}
    R_{\mathrm{coh}} = \frac{2A^2(d_s+d_i)^2}{V_s + V_i+2C}.
    \label{eq:lda_R_coherent}
\end{equation}
For a lossless amplifier, writing $G=\cosh^2r$ gives $d_s+d_i=e^r$, and therefore
\begin{equation}
    R_{\mathrm{coh}} = \frac{2A^2e^{2r}}{e^{2r} + N_{\mathrm{sys},s} + N_{\mathrm{sys},i}},
    \label{eq:lda_R_coherent_lossless}
\end{equation}
which is the coherent-combination result of Ref.~\cite{Abdo2026enhancing} expressed in our centroid-separation convention.

For the same Gaussian model, 4D LDA optimizes over the full linear projection rather than restricting the calibrated signal and idler to equal amplitudes. Defining $B_s=N_{\mathrm{sys},s}+1/2$ and $B_i=N_{\mathrm{sys},i}+1/2$, the optimal direction in the $(I_s,I_i)$ subspace is
\begin{equation}
    w_{\mathrm{LDA}}^{(I)} \propto
    \begin{pmatrix}
        d_s B_i\\
        d_i(B_s-\eta_s)
    \end{pmatrix}.
    \label{eq:lda_weights_loss}
\end{equation}
Thus, whereas the coherent phase sweep adjusts the orientation and relative sign of the idler contribution, LDA can additionally optimize its amplitude relative to the signal. Since the aligned equal-amplitude coherent projection is one admissible linear projection, $R_{4\mathrm{D}}\geq R_{\mathrm{coh}}$, with
\begin{equation}
    R_{4\mathrm{D}}-R_{\mathrm{coh}} = \frac{2A^2\left[d_sB_i-d_i(B_s-\eta_s)\right]^2}{(V_sV_i-C^2)(V_s+V_i+2C)}.
    \label{eq:lda_coh_exact_difference}
\end{equation}

The distinction is particularly clear near unity gain. At $G=1$, $d_i=C=0$, so the idler carries no state-dependent information and LDA assigns it zero weight,
\begin{equation}
    R_{4\mathrm{D}}(1) = R_s(1) = \frac{2A^2\eta_s}{B_s},
    \qquad
    R_{\mathrm{coh}}(1) = \frac{2A^2\eta_s}{B_s+B_i}.
    \label{eq:lda_coh_unity_gain}
\end{equation}
A phase rotation cannot remove the idler noise in this limit because it preserves the magnitude of the idler weight.

At large gain, the two methods become equivalent for any nonideal chain with finite loss or nonzero downstream added noise,
\begin{equation}
    \lim_{G\rightarrow\infty}R_{\mathrm{coh}} = \lim_{G\rightarrow\infty}R_{4\mathrm{D}} = 2A^2.
    \label{eq:lda_coh_high_gain_equivalence}
\end{equation}
The ideal lossless and noiseless case is an exception. For $\eta_s=\eta_i=1$ and $N_{\mathrm{sys},s}=N_{\mathrm{sys},i}=0$, the exact results are
\begin{equation}
    R_{4\mathrm{D}}=4A^2,
    \qquad
    R_{\mathrm{coh}}=2A^2,
\end{equation}
for all gains. More generally, when the downstream noise and loss are small, convergence to the common high-gain limit can be slow, and the LDA advantage can remain appreciable at practical gain. For example, for $\eta_s=\eta_i=1$, $N_{\mathrm{sys},s}=N_{\mathrm{sys},i}=0.1$, and $G=3$, the model gives $R_{4\mathrm{D}}\simeq2.63A^2$, $R_s\simeq2.31A^2$, and $R_{\mathrm{coh}}\simeq1.96A^2$. Thus, the difference between the prescribed coherent combination and LDA is not the ability to optimize relative phase, but the additional freedom to optimize the relative signal--idler amplitude.\\

\subsection{Experimental comparison}
 
\begin{figure}
    \centering
    \includegraphics[width=\linewidth]{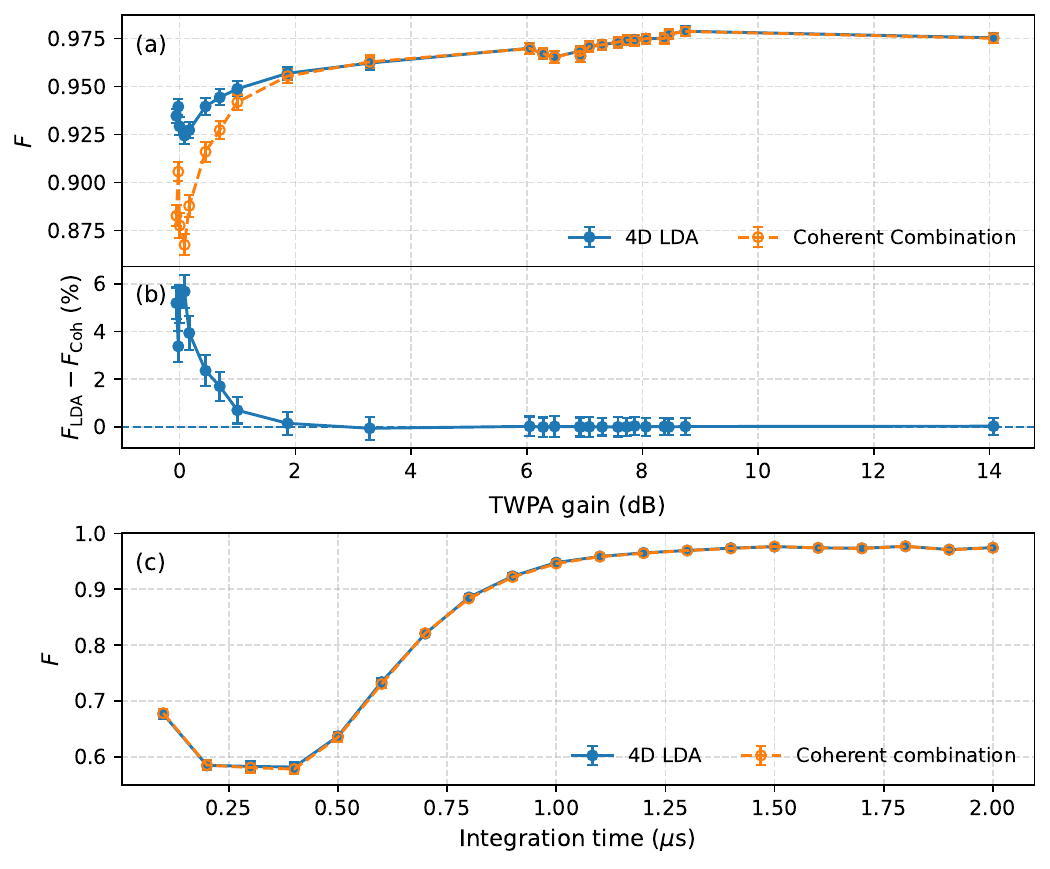}
    \caption{Experimental LDA vs. coherent-combination comparison. (a) Held-out assignment fidelity versus measured TWPA gain. (b) Paired fidelity difference $F_{\mathrm{4D}}-F_{\mathrm{coh}}$. (c) Fidelity versus integration time at a fixed higher-gain operating point. The coherent combination uses calibrated, independently aligned signal and idler quadratures with equal amplitude weights and a training-selected relative sign.}
    \label{fig:lda_vs_coherent_fidelity}
\end{figure}

We compare the two methods using the same simultaneously acquired signal and idler shots. For the coherent baseline, both channels are calibrated to common units and their state-dependent displacement axes are aligned independently. The relative idler sign is selected from $0$ and $\pi$ using the training data, after which the aligned quadratures are added with equal amplitude weights. This implements the aligned equal-amplitude coherent combination considered above, but does not constitute a continuous optimization over arbitrary relative amplitude and phase. For LDA, the four-dimensional projection is learned directly from the training-set centroids and pooled covariance.

Figures~\ref{fig:lda_vs_coherent_fidelity}(a) and (b) compare the two methods as a function of measured TWPA gain. The largest mean paired difference, $5.16\,\%$, occurs at the near-unity-gain point $0.08\,\mathrm{dB}$, where the coherent combination retains idler noise even though the idler displacement is small. Across the measured gain points at and above $2\,\mathrm{dB}$, the mean absolute pointwise fidelity difference is $0.01$ percentage points. Figure~\ref{fig:lda_vs_coherent_fidelity}(c) compares the methods versus integration time at the higher-gain operating point used for the main sweep; the mean absolute pointwise fidelity difference is $0.10\,\%$. Thus, for the noise levels and gains accessed in our experiment, the implemented coherent combination analysis and 4D LDA give closely matched performance away from gain values near unity.

\bibliography{apssamp}

\end{document}